# Gravitational Waves and Gamma-Rays from a Binary Neutron Star Merger: GW170817 and GRB 170817A

LIGO Scientific Collaboration and Virgo Collaboration, *Fermi* Gamma-ray Burst Monitor, and INTEGRAL
(See the end matter for the full list of authors.)

## Abstract

On 2017 August 17, the gravitational-wave event GW170817 was observed by the Advanced LIGO and Virgo detectors, and the gamma-ray burst (GRB) GRB 170817A was observed independently by the *Fermi* Gamma-ray Burst Monitor, and the Anti-Coincidence Shield for the Spectrometer for the *International Gamma-Ray Astrophysics Laboratory*. The probability of the near-simultaneous temporal and spatial observation of GRB 170817A and GW170817 occurring by chance is $5.0 \times 10^{-8}$. We therefore confirm binary neutron star mergers as a progenitor of short GRBs. The association of GW170817 and GRB 170817A provides new insight into fundamental physics and the origin of short GRBs. We use the observed time delay of $(+1.74 \pm 0.05)$ s between GRB 170817A and GW170817 to: (i) constrain the difference between the speed of gravity and the speed of light to be between $-3 \times 10^{-15}$ and $+7 \times 10^{-16}$ times the speed of light, (ii) place new bounds on the violation of Lorentz invariance, (iii) present a new test of the equivalence principle by constraining the Shapiro delay between gravitational and electromagnetic radiation. We also use the time delay to constrain the size and bulk Lorentz factor of the region emitting the gamma-rays. GRB 170817A is the closest short GRB with a known distance, but is between 2 and 6 orders of magnitude less energetic than other bursts with measured redshift. A new generation of gamma-ray detectors, and subthreshold searches in existing detectors, will be essential to detect similar short bursts at greater distances. Finally, we predict a joint detection rate for the *Fermi* Gamma-ray Burst Monitor and the Advanced LIGO and Virgo detectors of 0.1–1.4 per year during the 2018–2019 observing run and 0.3–1.7 per year at design sensitivity.

## 1. Introduction and Background

GW170817 and GRB 170817A mark the discovery of a binary neutron star (BNS) merger detected both as a gravitational wave (GW; LIGO Scientific Collaboration & Virgo Collaboration 2017a) and a short-duration gamma-ray burst (SGRB; Goldstein et al. 2017; Savchenko et al. 2017b). Detecting GW radiation from the coalescence of BNS and neutron star (NS)–black hole (BH) binary systems has been a major goal (Abbott et al. 2017a) of the LIGO (Aasi et al. 2015) and Virgo (Acernese et al. 2015) experiments. This was at least partly motivated by their promise of being the most likely sources of simultaneously detectable GW and electromagnetic (EM) radiation from the same source. This is important as joint detections enable a wealth of science unavailable from either messenger alone (Abbott et al. 2017f). BNS mergers are predicted to yield signatures across the EM spectrum (Metzger & Berger 2012; Piran et al. 2013), including SGRBs (Blinnikov et al. 1984; Paczynski 1986; Eichler et al. 1989; Paczynski 1991; Narayan et al. 1992), which produce prompt emission in gamma-rays and longer-lived afterglows.

A major astrophysical implication of a joint detection of an SGRB and of GWs from a BNS merger is the confirmation that these binaries are indeed the progenitors of at least some SGRBs. GRBs are classified as short or long depending on the duration of their prompt gamma-ray emission. This cut is based on spectral differences in gamma-rays and the bimodality of the observed distribution of these durations (Dezalay et al. 1992; Kouveliotou et al. 1993). This empirical division was accompanied by hypotheses that the two classes have different progenitors. Long GRBs have been firmly connected to the collapse of massive stars through the detection of associated Type Ibc core-collapse supernovae (see Galama et al. 1998, as well as Hjorth & Bloom 2012 and references therein). Prior to the results reported here, support for the connection between SGRBs and mergers of BNSs (or NS–BH binaries) came only from indirect observational evidence (Nakar 2007; Berger et al. 2013; Tanvir et al. 2013; Berger 2014), population synthesis studies (Bloom et al. 1999; Fryer et al. 1999; Belczynski et al. 2006), and numerical simulations (e.g., Aloy et al. 2005; Rezzolla et al. 2011; Kiuchi et al. 2015; Baiotti & Rezzolla 2017; Kawamura et al. 2016; Ruiz et al. 2016). The unambiguous joint detection of GW and EM radiation from the same event confirms that BNS mergers are progenitors of (at least some) SGRBs.

In Section 2 we describe the independent observations of GW170817 by the LIGO–Virgo and of GRB 170817A by the *Fermi* Gamma-ray Burst Monitor (GBM) and by the SPectrometer on board *INTEGRAL* Anti-Coincidence Shield (SPI-ACS). In Section 3 we establish the firm association between GW170817 and GRB 170817A. In Section 4 we explore the constraints on fundamental physics that can be obtained from the time separation between the GW and EM signals. In Section 5 we explore the implications of the joint detection of GW170817 and GRB 170817A on the SGRB engine and the NS equation of state (EOS). In Section 6 we explore the implications of the comparative dimness of GRB 170817A relative to the known SGRB population and revise the expectation rates for joint BNS–SGRB detections in the light of this discovery.

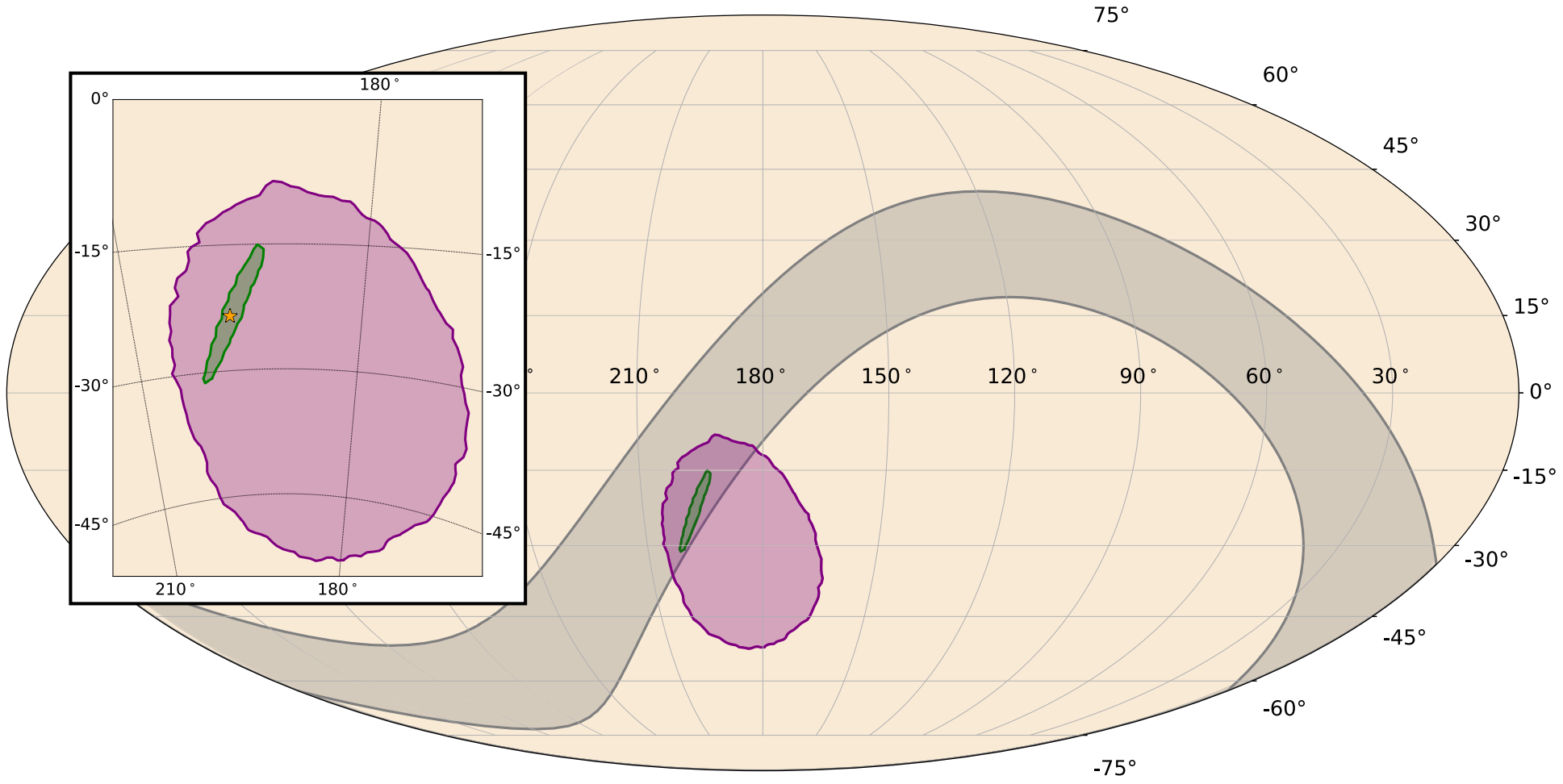

**Figure 1.** Final localizations. The 90% contour for the final sky-localization map from LIGO–Virgo is shown in green (LIGO Scientific Collaboration & Virgo Collaboration 2017a, 2017b, 2017c). The 90% GBM targeted search localization is overlaid in purple (Goldstein et al. 2017). The 90% annulus determined with *Fermi* and *INTEGRAL* timing information is shaded in gray (Svinkin et al. 2017). The zoomed inset also shows the position of the optical transient marked as a yellow star (Abbott et al. 2017f; Coulter et al. 2017a, 2017b). The axes are R.A. and decl. in the Equatorial coordinate system.

## 2. Observational Results

The observations of GW170817 and of GRB 170817A are described in detail in Abbott et al. (2017e), Goldstein et al. (2017), and Savchenko et al. (2017b). Here we summarize the observations relevant to the results presented in this Letter and report the results of two fully coherent searches for GWs from the sky location of GRB 170817A. For convenience, all measurements of time have been converted to their geocentric equivalent.

### 2.1. *LIGO–Virgo Observation of GW170817*

GW170817 is a GW signal from the inspiral of two low-mass compact objects and is the first GW observation consistent with a BNS coalescence (Abbott et al. 2017e, 2017f). GW170817 was first observed by a low-latency search (Cannon et al. 2012; Messick et al. 2017) on 2017 August 17 at 12:41:04 UTC as a single-detector trigger in the LIGO-Hanford detector (Abbott et al. 2017e; LIGO Scientific Collaboration & Virgo Collaboration 2017a). The temporal proximity of GRB 170817A was immediately identified by automatic comparison of the *Fermi*-GBM Gamma-ray Coordinates Network notice to the GW trigger (Urban 2016). Rapid offline re-analysis (Usman et al. 2016; Nitz et al. 2017b) of data from the LIGO/Virgo network confirmed the presence of a significant coincident signal in the LIGO GW detectors with a combined signal-to-noise ratio (S/N) of 32.4. The combination of observations from the LIGO and Virgo detectors allowed a precise sky position localization to an area of 28 deg$^2$ at 90% probability shown in green in Figure 1 (Abbott et al. 2017e; LIGO Scientific Collaboration & Virgo Collaboration 2017b). A time-frequency representation of the LIGO data containing GW170817 is shown in the bottom panel of Figure 2. The GPS time of the merger of GW170817 is $T_0^{\mathrm{GW}} = 1187008882.430^{+0.002}_{-0.002}$ s (Abbott et al. 2017e). At the observed signal strength, the false alarm rate of the all-sky search for compact-object mergers is less than 1 in 80,000 years (Abbott et al. 2017e). The offline searches target binaries with (detector frame) total mass 2–500 $M_{\odot}$. Signals are required to be coincident in time and mass in the LIGO detectors, but Virgo data are not used in the significance estimates of the all-sky offline search (Abbott et al. 2017e).

We present the results of two offline targeted searches that coherently combine the data from the LIGO and Virgo detectors and restrict the signal offset time and sky-location using information from the EM observation of GRB 170817A. The onset of gamma-ray emission from a BNS merger progenitor is predicted to be within a few seconds after the merger, given that the central engine is expected to form within a few seconds and that the jet propagation delays are at most of the order of the SGRB duration (see, e.g., Finn et al. 1999; Abadie et al. 2012 and references therein). The gravitational and EM waves are expected to travel at the same speed.

The first targeted search (Harry & Fairhurst 2011; Williamson et al. 2014; Abbott et al. 2017b; Nitz et al. 2017a) assumes that the source is a BNS or NS–BH binary merger and is located at the sky-position observed for the optical counterpart to GW170817 and GRB 170817A (Coulter et al. 2017a, 2017b; Abbott et al. 2017f) and that there is a $[-1, +5]$ s time delay in the arrival of gamma-rays (determined by the GBM trigger time) compared to the binary merger time (Abbott et al. 2017b). At the detection statistic value assigned to GW170817, this search has a $p$-value of $<9.4 \times 10^{-6}$($>4.2\sigma$), with this significance estimate limited by computational resources used to estimate the noise background. The second coherent search does not assume any particular GW morphology or GRB model (Sutton et al. 2010; Was et al. 2012; Abbott et al. 2017b) and uses the GBM localization of GRB 170817A to constrain the sky location of the source. This search allows for a $[-60, +600]$ s coincidence between the gamma-rays and the GWs in order to include potentially larger delays in collapsar models of long GRBs. At the detection-statistic value observed for GW170817, this search has a $p$-value of $1.3 \times 10^{-5}$ ($4.2\sigma$).

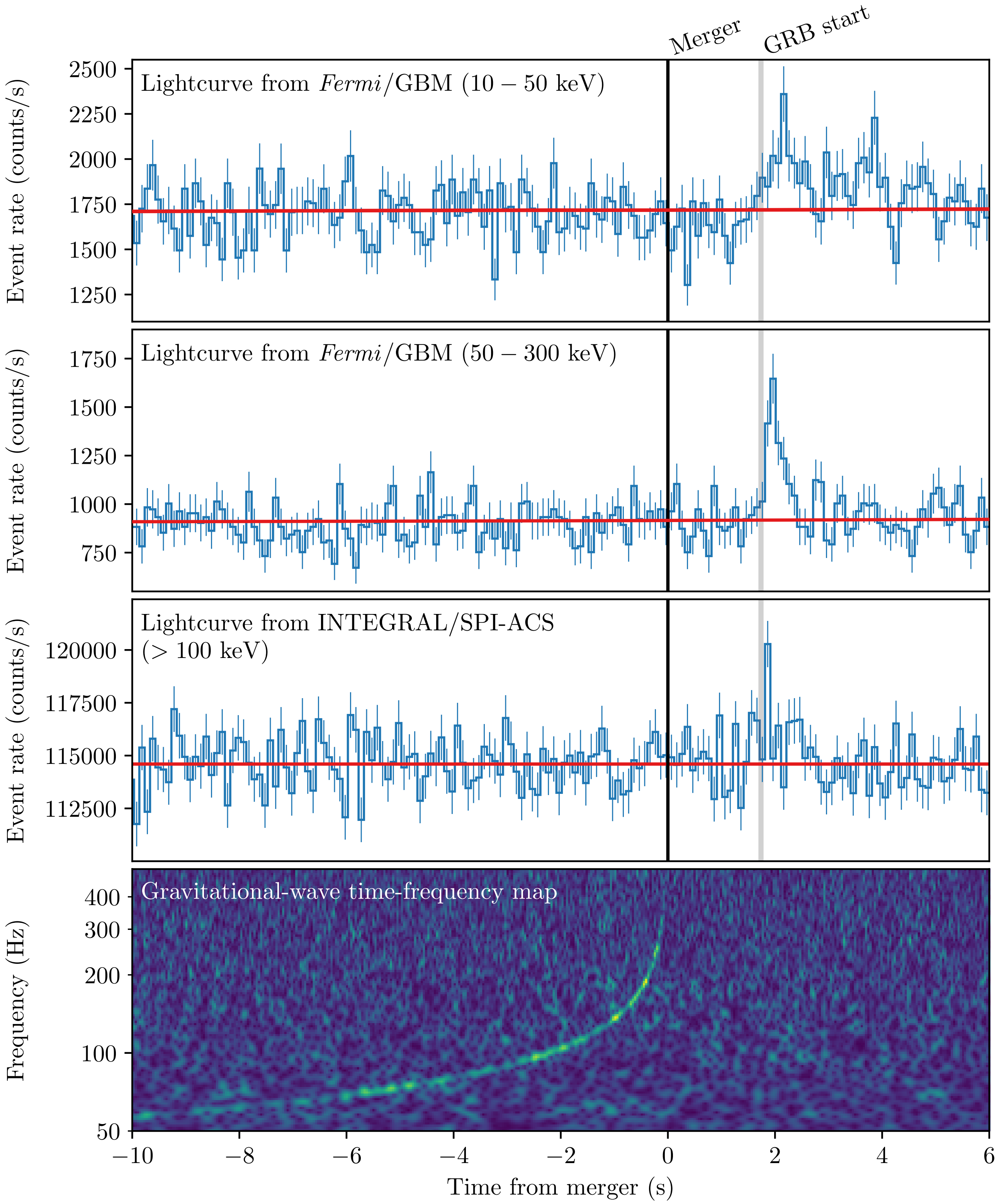

**Figure 2.** Joint, multi-messenger detection of GW170817 and GRB 170817A. Top: the summed GBM lightcurve for sodium iodide (NaI) detectors 1, 2, and 5 for GRB 170817A between 10 and 50 keV, matching the 100 ms time bins of the SPI-ACS data. The background estimate from Goldstein et al. (2016) is overlaid in red. Second: the same as the top panel but in the 50–300 keV energy range. Third: the SPI-ACS lightcurve with the energy range starting approximately at 100 keV and with a high energy limit of least 80 MeV. Bottom: the time-frequency map of GW170817 was obtained by coherently combining LIGO-Hanford and LIGO-Livingston data. All times here are referenced to the GW170817 trigger time $T_0^{\rm GW}$.

The 90% credible intervals (Veitch et al. 2015; Abbott et al. 2017e) for the component masses (in the $m_1 \geqslant m_2$ convention) are $m_1 \in (1.36,\ 2.26)\ M_\odot$ and $m_2 \in (0.86,\ 1.36)\ M_\odot$, with total mass $2.82^{+0.47}_{-0.09}\ M_\odot$, when considering dimensionless spins with magnitudes up to 0.89 (high-spin prior, hereafter). When the dimensionless spin prior is restricted to $\leqslant$0.05 (low-spin prior, hereafter), the measured component masses are $m_1 \in (1.36,\ 1.60)\ M_\odot$ and $m_2 \in (1.17,\ 1.36)\ M_\odot$, and the total mass is $2.74^{+0.04}_{-0.01}\,M_\odot$. This result is consistent with the masses of all known BNS systems (Ozel & Freire 2016; Tauris et al. 2017). From the GW signal, the best measured combination of the masses is the chirp mass $\mathcal{M} = (m_1 m_2)^{3/5}/(m_1 + m_2)^{1/5}$, which in the detector frame is found to be $1.1977^{+0.0008}_{-0.0003}\,M_\odot$.

The detection of GW170817 triggered a campaign of EM follow-up observations which led to the identification of NGC 4993 as the host galaxy of GW170817/GRB 170817A (Coulter et al. 2017a, 2017b; Abbott et al. 2017f). We evaluate the distance to the host galaxy from the ratio of the Hubble flow velocity of the host $3017 \pm 166\,\mathrm{km\,s^{-1}}$ (Abbott et al. 2017g) and two current measurements of the Hubble constant (Ade et al. 2016; Riess et al. 2016). These two distance measures are within a combined range of $(42.9 \pm 3.2)$ Mpc, which is consistent with the distance of $40^{+8}_{-14}$ Mpc determined with GW data alone and makes GW170817 the closest GW event ever observed (Abbott et al. 2016a, 2016b, 2017c, 2017d, 2017e).

The GW data constrain the inclination angle $\theta_{JN}$ between the total angular momentum of the system and the line of sight to be anti-aligned, with $\cos\theta_{JN} \leqslant -0.54$ (Abbott et al. 2017e). As the binary system component masses are comparable, the NS spins have little impact on the total angular momentum which is aligned with the orbital angular momentum within a few degrees. For discussions in this Letter we will assume that the orbital and total angular momenta are aligned. The SGRB jet is expected to be perpendicular to the accretion disk of the central engine if powered by neutrino annihilation or aligned with the magnetic pole of the rotating central object (Shibata et al. 2006), hence we assume the SGRB jet is aligned with the system rotation axis. This yields a jet viewing angle $\zeta = \min(\theta_{JN}, 180^\circ - \theta_{JN}) \leqslant 56^\circ$. As the distance measurement is correlated with $\theta_{JN}$, the known distance to NGC 4993 further constrains the viewing angle to $\zeta \leqslant 36^\circ$ or $\zeta \leqslant 28^\circ$ depending on the assumed value of the Hubble constant (Abbott et al. 2017g), with smaller values of the Hubble constant giving smaller misalignment angles.

### 2.2. Fermi-GBM Observation of GRB 170817A

GRB 170817A was autonomously detected in-orbit by the GBM flight software in a process known as "triggering." Goldstein et al. (2017) showed the signal exceeds $5\sigma$ in three (of twelve) GBM NaI detectors. The GBM detection showed two apparently distinct components. The triggering pulse, that lasts about half a second and falls within the usual observer distributions for GBM SGRBs, is shorter and spectrally harder than the subsequent softer, weaker emission that lasts a few seconds (Goldstein et al. 2017). Summed GBM lightcurves from the relevant detectors in two energy ranges, selected to show the two distinct components, are shown in the top two panels in Figure 2. The GBM time-tagged event data is binned to match the SPI-ACS temporal resolution (100 ms) and phase (matching bin edges) to allow for an easier comparison between the gamma-ray instruments.

Goldstein et al. (2017) quantify the likelihood of GRB 170817A being an SGRB based only on gamma-ray data. This is done by comparing the measured gamma-ray properties of GRB 170817A to the known distributions of short and long GRBs. Both the duration distribution alone and the duration and spectral hardness distributions together show that GRB 170817A is three times more likely to be an SGRB than a long GRB. These analyses are performed in a standard manner, resulting in a longer duration measure than apparent from the hard spike alone because the softer, weaker tail contributes to the calculated duration.

The final GBM localization of GRB 170817A (including systematic error) calculated by the GBM targeted search pipeline is shown in Figure 1. This pipeline performs a coherent search over all GBM detectors (NaI and BGO) and was originally developed to find gamma-ray signals below the onboard triggering threshold around GW triggers (Blackburn et al. 2015; Connaughton et al. 2016; Goldstein et al. 2016). The 50% and 90% credible regions cover $\sim 350\,\mathrm{deg}^2$ and $\sim 1100\,\mathrm{deg}^2$, respectively.

Fitting the main pulse in the GBM data with a parameterized function commonly used for GRB pulses indicates a gamma-ray emission onset of $0.310 \pm 0.048$ s before $T_0^{\mathrm{GBM}}$, where $T_0^{\mathrm{GBM}}$ is defined as the time of the GBM trigger (Goldstein et al. 2017). Based on the position of the optical transient, the signal arrives at *Fermi* 3.176 ms before it arrives at geocenter. With this correction we find that the start of the gamma-ray emission relative to the $T_0^{\mathrm{GW}}$ is $(+1.74 \pm 0.05)$ s. In this Letter all derived gamma-ray results use 68% confidence levels.

The spectral analysis using the standard GBM catalog criteria uses data from the 256 ms time interval between $T_0^{\mathrm{GBM}} - 0.192$ s and $T_0^{\mathrm{GBM}} + 0.064$ s. A fit to the "Comptonized" function, a power law with a high-energy exponential cutoff (see Goldstein et al. 2017 for a detailed explanation of this function), is preferred over both a simple power-law fit or models with more parameters. The fit produces values of $E_{\mathrm{peak}} = (215 \pm 54)$ keV, and a poorly constrained power-law index $\alpha = 0.14 \pm 0.59$. The average flux for this interval in the 10–1000 keV range is $(5.5 \pm 1.2) \times 10^{-7}\,\mathrm{erg\,s^{-1}\,cm^{-2}}$ with a corresponding fluence of $(1.4 \pm 0.3) \times 10^{-7}\,\mathrm{erg\,cm^{-2}}$. The shorter peak interval selection from $T_0^{\mathrm{GBM}} - 0.128$ s to $T_0^{\mathrm{GBM}} - 0.064$ s fit prefers the Comptonized function, yielding consistent parameters $E_{\mathrm{peak}} = (229 \pm 78)$ keV, $\alpha = 0.85 \pm 1.38$, and peak energy flux in the 10–1000 keV of $(7.3 \pm 2.5) \times 10^{-7}\,\mathrm{erg\,s^{-1}\,cm^{-2}}$. These standard fits are used to compare GRB 170817A to the rest of the SGRBs detected by GBM and to place GRB 170817A in context with the population of SGRBs with known redshift.

More detailed analysis included spectral fits to the two apparently distinct components. The main emission episode, represented by the peak in Figure 2, appears as a typical SGRB best fit by a power law with an exponential cutoff with spectral index $\alpha = -0.62 \pm 0.40$ and $E_{\mathrm{peak}} = (185 \pm 62)$ keV over a time interval $T_0^{\mathrm{GBM}} - 0.320$ s to $T_0^{\mathrm{GBM}} + 0.256$ s. The time-averaged flux is $(3.1 \pm 0.7) \times 10^{-7}\,\mathrm{erg\,s^{-1}\,cm^{-2}}$. The tail emission that appears spectrally soft is best fit by a blackbody (BB) spectrum, with temperature of $k_{\mathrm{B}}T = (10.3 \pm 1.5)$ keV and a time-averaged flux of $(0.53 \pm 0.10) \times 10^{-7}\,\mathrm{erg\,s^{-1}cm^{-2}}$, with selected source interval $T_0^{\mathrm{GBM}} + 0.832$ s to $T_0^{\mathrm{GBM}} + 1.984$ s. However, this emission is too weak and near the lower energy detection bound of GBM to completely rule out a non-thermal spectrum.

The temporal analysis yielded a $T_{90}$, defined as the time interval over which 90% of the burst fluence between 50–300 keV is accumulated, of $(2.0 \pm 0.5)$ s starting at $T_0^{\mathrm{GBM}} - 0.192$ s. The duration extends beyond the main emission pulse due to the soft component. This analysis reports a 64 ms peak photon flux of $(3.7 \pm 0.9)$ photons $\mathrm{s^{-1}\,cm^{-2}}$ and occurs from $T_0^{\mathrm{GBM}} + 0.0$ s to $T_0^{\mathrm{GBM}} + 0.064$ s. The minimum variability timescale for GRB 170817A is (0.125 ± 0.064) s.

Using the soft spectral template of the GBM targeted search, a Band function (Band et al. 1993) with a low energy power law index of −1.9, a high energy index of −3.7, and an $E_{\rm peak}$ of 70 keV, Goldstein et al. (2017) also set $3\sigma$ flux upper limits on precursor impulsive gamma-ray emission. The limits on precursor activity out to $T_0^{\rm GBM} - 200$ s are $(6.8\text{–}7.3) \times 10^{-7}$ erg s$^{-1}$ cm$^{-2}$ and $(2.0\text{–}2.1) \times 10^{-7}$ erg s$^{-1}$ cm$^{-2}$ for signals of 0.1 s and 1.0 s duration, respectively. The tail emission of GRB 170817A is not consistent with the general behavior of SGRBs with extended emission (Kaneko et al. 2015). We set limits on possible extended emission over 10 s intervals out to $\sim T_0^{\rm GBM} + 100$ s is $(6.4\text{–}6.6) \times 10^{-8}$ erg s$^{-1}$ cm$^{-2}$. Additional upper limits for representative normal and harder spectra are provided in Goldstein et al. (2017) and are up to a factor of a few less constraining.

### 2.3. *INTEGRAL SPI-ACS Observation of GRB 170817A*

The orientation of *INTErnational Gamma-Ray Astrophysics Laboratory* (*INTEGRAL*) with respect to the LIGO–Virgo localization of GW170817 favored the observation by SPI-ACS and was such that the sensitivity of the Imager on Board the *INTEGRAL* Satellite (IBIS) was much lower in comparison (Savchenko et al. 2017b). For comparison of relative sensitivities of different *INTEGRAL* instruments see Savchenko et al. (2017a).

A routine follow-up search for short transients in SPI-ACS identified a single excess at $T_0^{\rm ACS} = T_0^{\rm GW} + 1.88$ s with $S/N = 4.6$ at the 0.1 s timescale. The correction to the geocentric system assumes the location of the optical transient and results in delay of the signal arrival to *INTEGRAL* of 148.96 ms. In order to compare the intensity of the event observed by SPI-ACS to the GBM measurement, we compute the range of fluences compatible with the SPI-ACS data in the [−0.320 s, +0.256 s] time interval centered in $T_0^{\rm GBM}$, assuming the GBM best fit spectral model in the same interval. We derive a fluence estimate of $(1.4 \pm 0.4) \times 10^{-7}$ erg cm$^{-2}$ (statistical uncertainty only) in the 75–2000 keV energy range, consistent with GBM.

The significance of the association between the GBM observation of GRB 170817A and the event observed by SPI-ACS is $4.2\sigma$. While SPI-ACS would not have alone reported this event as a GRB, it would have reported the event while searching around GW170817, with an independent association significance of $3.2\sigma$ (Savchenko et al. 2017b). SGRBs are routinely jointly detected by GBM and SPI-ACS and the association evidence from time coincidence (quoted above) as well as the consistency between the event fluences and temporal properties observed by the two instruments proves that both GBM and SPI-ACS observed the same event. The difference between the time of arrival of the signal in the SPI-ACS and GBM detectors can be exploited to improve the gamma-ray localization of GRB 170817A, which may be beneficial in future joint detections.

The significant interval of the SPI-ACS lightcurve of GRB 170817A is limited to a single pulse with a duration of 100 ms (third panel in Figure 1). GBM and SPI-ACS see the main pulse as appearing to have different durations because they are sensitive in different energy ranges. If the GBM data are shown in an energy range higher than the standard 50-300 keV, the main pulse is consistent with the 100 ms interval seen in SPI-ACS. The lightcurve observed by SPI-ACS reveals a short rise time (<50 ms) and a rapid drop (<50 ms). We therefore constrain the pulse duration in the energy range observed by SPI-ACS (∼75–2000 keV) to less than 100 ms.

## 3. Unambiguous Association

The separation of GRBs into short and long classes was suggested by their duration distributions and reinforced by differences in the prompt gamma-ray emission of the two classes (Dezalay et al. 1992; Kouveliotou et al. 1993). Tying the short class to a different progenitor from the long class was strengthened by redshift measurements of their hosts (Berger 2014). Association of SGRBs with older stellar populations than long GRBs was supported by the types of galaxies that host them (Fong et al. 2013); the connection to BNS mergers was strengthened by the offsets of SGRBs afterglows from their host galaxies (Troja et al. 2008; Church et al. 2011; Fong & Berger 2013) and by the absence of supernovae following nearby, well-observed SGRBs (Fox et al. 2005; Hjorth et al. 2005a, 2005b; Bloom et al. 2006; Soderberg et al. 2006; D'Avanzo et al. 2009; Kocevski et al. 2010; Rowlinson et al. 2010; Berger et al. 2013). We provide conclusive evidence for the BNS-SGRB connection by quantifying the chance temporal and spatial coincidence for GRB 170817A and GW170817 arising from two independent astrophysical events.

To quantify the temporal agreement, we consider the null hypothesis that SGRB and GW detection events are independent Poisson processes and determine how unlikely it is to observe an unassociated SGRB within $\Delta t_{\rm SGRB-GW} = 1.74$ s of the GW signal. GWs from a BNS merger have been detected once to date, so the $p$-value is $P_{\rm temporal} = 2\Delta t_{\rm SGRB-GW} R_{\rm GBM-SGRB}$, where $R_{\rm GBM-SGRB}$ is the GBM SGRB detection rate. Using the standard duration cut $T_{90} < 2$ s, GBM triggered on-board in response to 351 SGRBs in 3324 days[184] (the number of days between the GBM on-board trigger activation and the detection of GRB 170817A). Further, we account for the livetime of GBM, which is disabled 15% of the time to preserve detector lifetime in regions of high particle activity during transit through the South Atlantic Anomaly. Therefore, $P_{\rm temporal} = 2(1.74\ {\rm s})(351/3324\ {\rm days}/0.85) = 5.0 \times 10^{-6}$, which corresponds to a $4.4\sigma$ significance in Gaussian statistics.

In order to quantify the spatial agreement of the independent GBM and LIGO–Virgo localizations, we define the statistic $\mathcal{S} = \sum_{i=1}^{N_{\rm pix}} P_{1i} P_{2i}$, where $P_1$ and $P_2$ are the posterior probabilities from GBM and LIGO–Virgo maps and $i$ is the HEALPix (Gorski et al. 2005) pixel index. $\mathcal{S}$ is then compared against a background distribution generated by randomly shifting and rotating GBM posteriors from a representative sample of 164 SGRBs localized by the targeted search. We factor in the estimated localization systematic, and randomly shift and rotate each map 10 times. This background method accounts for the morphology and size distributions of GBM SGRB localizations. We find a $p$-value $P_{\rm spatial} = 0.01$ that the two independent localizations agree this well by chance.

The temporal and spatial $p$-values are independent quantities, thus the probability that GRB 170817A and GW170817 occurred this close in time and with this level of location agreement by chance is $P_{\rm temporal} \times P_{\rm spatial} = (5.0 \times 10^{-6}) \times (0.01) = 5.0 \times 10^{-8}$, corresponding to a Gaussian-equivalent significance of $5.3\sigma$. This unambiguous association confirms that BNS mergers are progenitors of (at least some) SGRBs.

[184] https://heasarc.gsfc.nasa.gov/W3Browse/fermi/fermigbrst.html

## 4. Implications for Fundamental Physics

Little or no arrival delay between photons and GWs over cosmological distances is expected as the intrinsic emission times are similar and the propagation speeds of EM and GWs are thought to be identical. In this Section we discuss the implications on fundamental physics of the temporal offset of $(+1.74 \pm 0.05)$ s measured between GW170817 and GRB 170817A.

Standard EM theory minimally coupled to general relativity predicts that GWs and light propagate with identical speeds. The refractive index of vacuum is expected to be unity, and both waves are expected to be affected by background gravitational potentials in the same way. The arrival delay of only a few seconds across a distance greater than one hundred million light years places stringent constraints on deviations from fundamental principles. We use the observed temporal offset, the distance to the source, and the expected emission-time difference to place constraints on the deviation of the speed of gravity from the speed of light, and on violations of Lorentz invariance and the equivalence principle.

### *4.1. Speed of Gravity*

Assuming a small difference in travel time $\Delta t$ between photons and GWs, and the known travel distance $D$, the fractional speed difference during the trip can be written $\Delta v / v_{\rm EM} \approx v_{\rm EM} \Delta t / D$, where $\Delta v = v_{\rm GW} - v_{\rm EM}$ is the difference between the speed of gravity $v_{\rm GW}$ and the speed of light $v_{\rm EM}$. This relation is less constraining for small distances, hence we conservatively use here $D = 26$ Mpc, the lower bound of the 90% credible interval on luminosity distance derived from the GW signal (Abbott et al. 2017e). If we conservatively assume that the peak of the GW signal and the first photons were emitted simultaneously, attributing the entire $(+1.74 \pm 0.05)$ s lag to faster travel by the GW signal, this time difference provides an upper bound on $\Delta v$. To obtain a lower bound on $\Delta v$, one can assume that the two signals were emitted at times differing by more than $(+1.74 \pm 0.05)$ s with the faster EM signal making up some of the difference. As a conservative bound relative to the few second delays discussed in Section 2.1, we assume the SGRB signal was emitted 10 s after the GW signal. The resulting constraint on the fractional speed difference is

$$-3 \times 10^{-15} \leqslant \frac{\Delta v}{v_{\rm EM}} \leqslant +7 \times 10^{-16}. \qquad (1)$$

The intergalactic medium dispersion has negligible impact on the gamma-ray photon speed, with an expected propagation delay many orders of magnitude smaller than our errors on $v_{\rm GW}$.

Lags much longer than 10 s are proposed in alternative models (e.g., Ciolfi & Siegel 2015; Rezzolla & Kumar 2015), and emission of photons before the merger is also possible (Tsang et al. 2012). Hence, certain exotic scenarios can extend this time difference window to (−100 s, 1000 s), yielding a 2 orders of magnitude broadening of the allowed velocity range on either side. While the emission times of the two messengers are inherently model dependent, conservative assumptions yield dramatic improvements over existing indirect (Kostelecky & Russell 2017) and direct (Cornish et al. 2017) constraints, which allow for time differences of more than 1000 years. Future joint GW–GRB detection should allow disentangling the emission time difference from the relative propagation time, as only the latter is expected to depend on distance.

### *4.2. Lorentz Invariance Violation Limits*

Within a comprehensive effective field theory description of Lorentz violation (Colladay & Kostelecký 1997, 1998; Kostelecký 2004; Tasson 2014), the relative group velocity of GWs and EM waves, is controlled by differences in coefficients for Lorentz violation in the gravitational sector and the photon sector at each mass dimension $d$ (Kostelecký & Mewes 2016, 2009, 2008; Wei et al. 2017). We focus here on the non-birefringent, non-dispersive limit at mass dimension $d = 4$, as it yields by far the most impressive results. In this case, the difference in group velocities for the two sectors takes the form

$$\Delta v = -\sum_{\substack{\ell m \\ \ell \leqslant 2}} Y_{\ell m}(\hat{n}) \left( \frac{1}{2} (-1)^{1+\ell} \bar{s}^{(4)}_{\ell m} - c^{(4)}_{(I)\ell m} \right). \qquad (2)$$

The result is presented in a spherical harmonic, $Y_{\ell m}$, basis, $\bar{s}^{(4)}_{\ell m}$ and $c^{(4)}_{(I)\ell m}$ being spherical-basis coefficients for Lorentz violation in the gravitational and EM sectors, respectively. The direction $\hat{n}$ refers to the sky position (provided in Coulter et al. 2017a, 2017b).

For ease of comparison with the many existing sensitivities (Shao 2014a, 2014b; Shao et al. 2017; Kostelecký & Tasson 2015; Bourgoin et al. 2016; Le Poncin-Lafitte et al. 2016; Kostelecky & Russell 2017) to the $d = 4$ gravity-sector coefficients (Bailey & Kostelecký 2006; Hees et al. 2016), an analysis in which the coefficients are constrained one at a time is useful (Flowers et al. 2016), with all other coefficients, including the EM sector ones, set to zero. These results are presented in Table 1 along with the best constraints for each coefficient prior to this work. These results can be compared with the isotropic $A$, $\alpha_{\rm LV}$ Lorentz violation parametrization (Mirshekari et al. 2012) used by Abbott et al. (2017c) in dispersive GW tests. The $\alpha_{\rm LV} = 2$ limit of this parametrization is equivalent to the isotropic limit of the framework discussed above, with $\bar{s}^{(4)}_{00} \rightarrow \sqrt{4\pi} A$. Constraints on $A$ for $\alpha_{\rm LV} = 2$ can be obtained from the first line of Table 1; these cannot be established within the analysis carried out in Abbott et al. (2017c).

### *4.3. Test of the Equivalence Principle*

Probing whether EM radiation and GWs are affected by background gravitational potentials in the same way is a test of the equivalence principle (Will 2014). One way to achieve this is to use the Shapiro effect (Shapiro 1964), which predicts that the propagation time of massless particles in curved spacetime, i.e., through gravitational fields, is slightly increased with respect to the flat spacetime case. We will consider the following simple parametrized form of the Shapiro delay (Krauss & Tremaine 1988; Longo 1988; Gao et al. 2015; Kahya & Desai 2016):

$$\delta t_{\rm S} = -\frac{1+\gamma}{c^3} \int_{\boldsymbol{r}_{\rm e}}^{\boldsymbol{r}_{\rm o}} U(\boldsymbol{r}(l)) dl, \qquad (3)$$

**Table 1**
Constraints on the Dimensionless Minimal Gravity Sector Coefficients

| $\ell$ | Previous Lower | This Work Lower | Coefficient | This Work Upper | Previous Upper |
|---|---|---|---|---|---|
| 0 | $-3 \times 10^{-14}$ | $-2 \times 10^{-14}$ | $\bar{s}_{00}^{(4)}$ | $5 \times 10^{-15}$ | $8 \times 10^{-5}$ |
| 1 | $-1 \times 10^{-13}$ | $-3 \times 10^{-14}$ | $\bar{s}_{10}^{(4)}$ | $7 \times 10^{-15}$ | $7 \times 10^{-14}$ |
| | $-8 \times 10^{-14}$ | $-1 \times 10^{-14}$ | $-\text{Re } \bar{s}_{11}^{(4)}$ | $2 \times 10^{-15}$ | $8 \times 10^{-14}$ |
| | $-7 \times 10^{-14}$ | $-3 \times 10^{-14}$ | $\text{Im } \bar{s}_{11}^{(4)}$ | $7 \times 10^{-15}$ | $9 \times 10^{-14}$ |
| 2 | $-1 \times 10^{-13}$ | $-4 \times 10^{-14}$ | $-\bar{s}_{20}^{(4)}$ | $8 \times 10^{-15}$ | $7 \times 10^{-14}$ |
| | $-7 \times 10^{-14}$ | $-1 \times 10^{-14}$ | $-\text{Re } \bar{s}_{21}^{(4)}$ | $2 \times 10^{-15}$ | $7 \times 10^{-14}$ |
| | $-5 \times 10^{-14}$ | $-4 \times 10^{-14}$ | $\text{Im } \bar{s}_{21}^{(4)}$ | $8 \times 10^{-15}$ | $8 \times 10^{-14}$ |
| | $-6 \times 10^{-14}$ | $-1 \times 10^{-14}$ | $\text{Re } \bar{s}_{22}^{(4)}$ | $3 \times 10^{-15}$ | $8 \times 10^{-14}$ |
| | $-7 \times 10^{-14}$ | $-2 \times 10^{-14}$ | $-\text{Im } \bar{s}_{22}^{(4)}$ | $4 \times 10^{-15}$ | $7 \times 10^{-14}$ |

**Note.** Constraints on the dimensionless minimal gravity sector coefficients obtained in this work via Equations (1) and (2) appear in columns 3 and 5. The corresponding limits that predate this work and are reported in columns 2 and 6; all pre-existing limits are taken from Kostelecký & Tasson (2015), with the exception of the upper limit on $\bar{s}_{00}^{(4)}$ from Shao (2014a, 2014b). The isotropic upper bound in the first line shows greater than 10 orders of magnitude improvement. The gravity sector coefficients are constrained one at a time, by setting all other coefficients, including those from the EM sector, to zero.

where $\boldsymbol{r}_{\mathrm{e}}$ and $\boldsymbol{r}_{\mathrm{o}}$ denote emission and observation positions, respectively, $U(\boldsymbol{r})$ is the gravitational potential, and the integral is computed along the wave path. $\gamma$ parametrizes a deviation from the Einstein–Maxwell theory, which minimally couples classical electromagnetism to general relativity. We allow for different values of $\gamma$ for the propagation of EM and GWs ($\gamma_{\mathrm{EM}}$ and $\gamma_{\mathrm{GW}}$, respectively, with $\gamma_{\mathrm{EM}} = \gamma_{\mathrm{GW}} = 1$ in the Einstein–Maxwell theory).

While obtaining the best bound on the difference between the Shapiro time delays requires modeling the potential $U(\boldsymbol{r})$ along the entire line of sight, we determine a conservative bound on $\gamma_{\mathrm{GW}} - \gamma_{\mathrm{EM}}$ by considering only the effect of the Milky Way outside a sphere of 100 kpc, and by using a Keplerian potential with a mass of $2.5 \times 10^{11} M_{\odot}$ (the lowest total mass within a sphere of radius 100 kpc quoted in Bland-Hawthorn & Gerhard 2016, from Gibbons et al. 2014, taking the 95% confidence lower bound) (Krauss & Tremaine 1988; Longo 1988; Gao et al. 2015). Using the same time bounds as Equation (1) we find

$$-2.6 \times 10^{-7} \leqslant \gamma_{\mathrm{GW}} - \gamma_{\mathrm{EM}} \leqslant 1.2 \times 10^{-6}. \tag{4}$$

The best absolute bound on $\gamma_{\mathrm{EM}}$ is $\gamma_{\mathrm{EM}} - 1 = (2.1 \pm 2.3) \times 10^{-5}$, from the measurement of the Shapiro delay (at radio wavelengths) with the Cassini spacecraft (Bertotti et al. 2003).

## 5. Astrophysical Implications

The joint GW–GRB detection provides us with unprecedented information about the central engine of SGRBs. The delay between the GW and the GRB trigger times allows us to examine some basic GRB physics. This delay could be intrinsic to the central engine, reflecting the time elapsed from the moment the binary components come into contact to the formation of a remnant BH and the resulting jet. This interpretation includes the case of a relatively long-lived massive NS remnant, which has been suggested to survive from seconds to minutes after merger (see Faber & Rasio 2012; Baiotti & Rezzolla 2017 and references therein). The delay could also be due to the propagation time of the relativistic jet, including the time it takes for the jet to break out of the dense gaseous environment produced by non-relativistic merger ejecta (Nagakura et al. 2014; Moharana & Piran 2017) and/or the emitting region to become transparent to gamma-rays (Mészáros & Rees 2000).

We first discuss the implications that the time delay between the GW and EM emission has on the physical properties of the emitting region when considering the jet propagation and transparency scenarios. Here we assume that the entire delay is due to the expansion of the emitting region and neglect any intrinsic delays between the moment of binary coalescence and the launching of the resulting jet, thus placing limits on the physical properties of the system. Then we consider the impact of SGRB emission from an NS merger on the EOS of dense matter.

### *5.1. GRB Physics*

The main hard peak observed for GRB 170817A lasted roughly half a second. This peak is consistent with a single intrinsic emission episode as it is well described by a single pulse (Goldstein et al. 2017), showing no evidence for significant substructure (spikes). This interpretation is consistent with the SPI-ACS observation of a single peak (Savchenko et al. 2017b). The GBM detection of GRB 170817A also shows no evidence for photons with energy >511 keV, implying that the outflow does not require a high bulk Lorentz factor $\Gamma$ to overcome photon–photon absorption at the source.

Explanations for the extreme energetics and short timescales observed in GRBs invoke a near instantaneous release of a large amount of energy in a compact volume of space (Goodman 1986; Paczynski 1986). This is commonly referred to as the fireball model, and it is the framework that we will assume for the remainder of this section. The fireball model is largely independent of the burst progenitor and focuses on the dynamics of such a system after this sudden release of energy. The resulting pair-plasma is optically thick and quickly expands under its own pressure to produce a highly relativistic outflow that coasts asymptotically with a constant Lorentz factor $\Gamma$. Within the fireball, kinetic energy is imparted to particles entrained in the outflow, although alternative models exist in which the energy outflow occurs mostly as Poynting flux (Usov 1992; Lyutikov & Blandford 2003). The observed gamma-ray pulses are attributed to shocks internal to this relativistic outflow, which convert some of their kinetic energy into the observed EM radiation (Rees & Meszaros 1994). These shocks could produce the predominantly non-thermal emission observed in most GRBs, although non-shock heating models have also been proposed (e.g., Giannios 2006). The overall multi-pulse duration of a burst is thought to reflect the time that the inner engine was active (e.g., producing inhomogeneities in the outflow represented as shells traveling with different bulk velocities) and the variability of individual pulses reflects the size of the shells producing the emission. For a top-hat jet model, $\Gamma$ is assumed constant over the jet surface and the observer never sees beyond the beaming angle $\theta_b \sim 1/\Gamma$. Therefore, the values inferred from the data are independent of the inclination angle from the total angular momentum axis of the system, as long as the viewer is within the opening angle of the jet.

We can examine the implications of the observed delay between the GW and EM signals in the internal shock scenario if we consider two shells emitted at time $t_{\rm GW} = 0$ and time $t_{\rm GW} + \Delta t_{\rm engine}$. If the Lorentz factor of the second shell, $\gamma_2$, is greater than the Lorentz factor of the first shell, $\gamma_1$, the shells will collide at time

$$t_{\rm delay} = \frac{\Delta t_{\rm engine}}{1 - (\gamma_1/\gamma_2)^2}, \quad (5)$$

which is valid if $\gamma_1, \gamma_2 \gg 1$. If the shells have comparable masses, conservation of energy and momentum leads to a merged shell with Lorentz factor $\gamma_{\rm m} = (\gamma_1\gamma_2)^{1/2}$. The resulting pulse profile is determined by two timescales. The rise time (which we equate to the minimum variability timescale) can be attributed to the light-crossing time of the individual emission regions and is expressed as

$$\Delta t_{\rm rise} \approx \frac{\delta R}{2c\gamma_{\rm m}^2}, \quad (6)$$

where $\delta R$ is the thickness of the emitting region. The decay time reflects angular effects, where off-axis emission is delayed and affected by a varying Doppler boost due to the curvature of the relativistic shell. This timescale is essentially the difference in light-travel time between photons emitted along the line of sight and photons emitted at an angle $\theta$ along a shell of radius $R$. This timescale may be expressed as

$$\Delta t_{\rm decay} = \frac{R(1 - \cos\Delta\theta)}{c} \approx \frac{R(\Delta\theta)^2}{2c} \approx \frac{R}{2c\gamma_{\rm m}^2} > \Delta t_{\rm rise}, \quad (7)$$

where we assume that the solid angle accessible to the observer is limited by relativistic beaming and thus given by $\theta \sim 1/\gamma$. At the same time, the distance that the first shell has traveled since ejection is $R_1 \approx 2c\gamma_1^2 t_{\rm delay}$, leading to

$$\Delta t_{\rm decay} \approx t_{\rm delay}(\gamma_1/\gamma_2) = \frac{\Delta t_{\rm engine}}{1 - (\gamma_1/\gamma_2)^2}\frac{\gamma_1}{\gamma_2}. \quad (8)$$

The conclusion is a linear correlation between the delay in the GW and EM signals and the resulting pulse duration, modulo the ratio of the Lorentz factors of the two colliding shells (Fenimore et al. 1996; Kocevski et al. 2007; Krimm et al. 2007).

The relative similarity between the gamma-ray duration $T_{90}$ and the delay between the GW and the EM emission gives $\Delta t_{\rm decay}/t_{\rm delay} \sim 1$, pointing to an internal shock scenario in which the difference in the Lorentz factors of the colliding shells, $\Delta\gamma$, is much smaller than their typical values, i.e., $\Delta\gamma \ll \gamma$. This would imply that the jet was launched shortly after the time of the merger and points to a relatively short $\Delta t_{\rm engine}$ time in which the central engine was active. Such a scenario would produce a collision that was relatively inefficient at converting the internal energy of the shocks to radiation, resulting in a significant isotropic equivalent kinetic energy remaining in the merged shell (Kobayashi et al. 1997; Krimm et al. 2007). This would lead to a very significant energy injection into the resulting afterglow, producing late time "refreshed shocks" (Rees & Mészáros 1998; Kumar & Piran 2000; Ramirez-Ruiz et al. 2001), which are typically not observed in the X-ray (and optical) lightcurves of SGRBs (e.g., Perley et al. 2009).

Some of these energetics constraints can be alleviated if we exclude the soft thermal emission from the gamma-ray duration estimate. In this case, the prompt non-thermal emission of $\Delta t_{\rm decay} \simeq 0.5$ s would be due to internal shocks and the soft thermal emission would be attributed to a separate component. In this case we obtain $\Delta t_{\rm decay}/t_{\rm delay} \simeq 0.3$, implying $\gamma_2 \approx 3\gamma_1$. These energetics considerations may suggest that the initial hard pulse and the subsequent thermal emission observed by GBM may indeed be distinct components.

Within the context of the internal shock model, if we assume the entire $(+1.74 \pm 0.05)$ s delay between the GW and the EM emission is due to jet propagation time and use a Lorentz factor of $\gamma < 100$ for the first shell, we can estimate an upper limit to the radius of the relativistic outflow to be $R \sim 5 \times 10^{14}$ cm or $\sim$30 au. The minimum variability timescale $\Delta t_{\rm rise} = \Delta t_{\rm min} \sim 0.125$ s (Goldstein et al. 2017) yields an upper limit on the size of the emitting region of $\delta R \sim 4 \times 10^{13}$ cm, or $\sim$3 au. The ratio of the two is independent of the unknown Lorentz factor and is of order $\delta R/R \sim 10\%$.

The single-pulsed nature of the gamma-ray emission, as well as the observed $\Delta t_{\rm decay}/t_{\rm delay} \sim 1$, also leaves open the possibility that the GBM signal is entirely of an external shock origin. In this scenario, the relativistic outflow converts its internal energy to radiation due to its interaction with an external medium, such as the interstellar matter (Meszaros & Rees 1992). If we associate the duration of the main pulse with the deceleration time, i.e., the timescale over which the jet is significantly decelerated by interstellar matter of constant density $n$, in the external shock scenario (Dermer et al. 1999):

$$t_{\rm dec} = [3E_{\rm k,iso}/(4\pi\gamma^8 n m_{\rm p} c^5)]^{1/3} = T_{\rm obs}. \quad (9)$$

$E_{\rm k,iso}$ is the kinetic energy of the jet calculated assuming a gamma-ray production efficiency of 20%, $m_{\rm p} = 1.67 \times 10^{-27}$ kg is the proton mass, $c$ is the speed of light, and $T_{\rm obs}$ is the approximate duration of the main peak. We can thus estimate the Lorentz factor of the jet in the external shock scenario to be

$$\gamma \approx 310\left(\frac{E_{\rm k,iso}}{2\times 10^{47}\ {\rm erg}}\right)^{1/8}\left(\frac{n}{0.1\ {\rm cm}^{-3}}\right)^{-1/8}\left(\frac{T_{\rm obs}}{0.5\ {\rm s}}\right)^{-3/8}. \quad (10)$$

The deceleration radius represents the upper limit of efficient energy extraction (even for internal shocks) and can be expressed as

$$R_{\rm dec} = 2\gamma^2 c T_{\rm obs} = 3.0\times 10^{15}\left(\frac{E_{\rm k,iso}}{2\times 10^{47}{\rm erg}}\right)^{1/4} \times \left(\frac{n}{0.1\ {\rm cm}^{-3}}\right)^{-1/4}\left(\frac{T_{\rm obs}}{0.5\ {\rm s}}\right)^{1/4}{\rm cm}. \quad (11)$$

Therefore, the deceleration radius and associated Lorentz factor also serve as upper limits to the radius and Lorentz factor of the emitting region in the internal shock scenario.

The soft thermal component observed by GBM could also be due to the photosphere of the fireball before it becomes optically thin to gamma-rays. In this interpretation, the delay between the GW and the GRB trigger times may represent the time it takes for the relativistic fireball to expand and become optically thin to gamma-ray radiation. We can examine this scenario by estimating the time it takes for a fireball to become transparent to high-energy radiation in an environment similar to that of a BNS merger.

Following Mészáros & Rees (2000, hereafter MR00), we assume an outflow with an initial radius $R_0 = 6GM_{\rm BH}/c^2$ (the innermost stable circular orbit of a Schwarzschild BH with a mass equal to $M_{\rm BH}$). In our case $R_0 = 2.5\times 10^6$ cm, and $M_{\rm BH} = 2.8\,M_\odot$. Given the GBM observations, we estimate an isotropic equivalent energy of the soft thermal BB component to be $E_{\rm iso,BB} = 1.3\times 10^{46}$ erg and peak isotropic luminosity of $L_{\rm iso,p} = 1.6\times 10^{47}\ {\rm erg\ s^{-1}}$ (see Section 6.1). We take this luminosity as an upper bound of the average luminosity, which may be estimated as $L_{\rm iso,BB} = E_{\rm iso,BB}/\Delta t_{\rm BB} = 1.1\times 10^{46}\ {\rm erg\ s^{-1}}$, where we have used a duration of the soft BB component of $\Delta t_{\rm BB} = 1.15$ s.

Using these parameters along with fiducial values (see Appendix B for details), we estimate the photosphere radius to be (MR00)

$$R_{\rm ph} = \frac{L\sigma_T Y}{4\pi m_{\rm p} c^3\eta^3} = 2.01\times 10^{13}\ {\rm cm} \times \left(\frac{L_0}{10^{50}\ {\rm erg\ s^{-1}}}\right)\left(\frac{\eta}{18}\right)^{-3} Y. \quad (12)$$

Where $L_0$ is the initial fireball luminosity, $Y$ is the number of electrons per baryon (in our case $Y \simeq 1$), $\sigma_T = 6.65\times 10^{-25}\ {\rm cm}^2$ is the Thomson cross-section, and $\eta$ is the dimensionless entropy of the fireball, whose value is much smaller than the canonical one in the standard fireball model (see Appendix B). We note that $L_0$ can be much larger than $L_{\rm iso,BB}$, since the fireball must expand and convert the remaining internal energy into kinetic energy of the ejecta.

The laboratory frame time needed for the fireball expanding at roughly the speed of light to reach the transparency radius is $t_{\rm ph} \simeq R_{\rm ph}/c$; thus,

$$t_{\rm ph} \simeq 672\ {\rm s}\left(\frac{L_0}{10^{50}\ {\rm erg\ s^{-1}}}\right)\left(\frac{\eta}{18}\right)^{-3} Y. \quad (13)$$

Following Bianco et al. (2001), we can set a upper bound to the conversion from the laboratory to observer frame by assuming the observer is viewing the fireball at most at an angle $\cos\vartheta = v/c$:

$$t_a \simeq \frac{t_{\rm ph}}{\gamma^2} \simeq 2.1\ {\rm s}\left(\frac{L_0}{10^{50}{\rm erg\ s^{-1}}}\right)\left(\frac{\eta}{18}\right)^{-5} Y. \quad (14)$$

This upper bound can account for the time delay between the GW and the prompt radiation in the soft thermal peak.

Employing MR00's Equation (8) we can estimate the observer frame temperature of the expanding fireball at the photospheric radius

$$T_{\rm ph}^{\rm obs} = \eta T_{\rm ph} \simeq 2.3\ {\rm keV}\left(\frac{L_0}{10^{50}{\rm erg\ s^{-1}}}\right)^{-5/12} \times \left(\frac{R_0}{2.5\times 10^6\ {\rm cm}}\right)^{-5/6}\left(\frac{\eta}{18}\right)^{11/3} Y^{-2/3}\gamma_0^{-5/6}. \quad (15)$$

$T_{\rm ph}^{\rm obs}$ can then be compared to the one obtained from observational fits to the GBM data, which provide a BB temperature of $T_{\rm BB}^{\rm obs} \simeq 10.3$ keV. Our result underestimates the observed BB temperature by a factor ∼4, but we are neglecting Comptonization effects, which may slightly raise the estimated temperature. The corresponding BB luminosity is (MR00)

$$L_{\rm ph} \simeq 1.7\times 10^{46}\ {\rm erg\ s^{-1}}\left(\frac{L_0}{10^{50}{\rm erg\ s^{-1}}}\right)^{7/12} \times \left(\frac{R_0}{2.5\times 10^6{\rm cm}}\right)^{-5/6}\left(\frac{\eta}{18}\right)^{8/3} Y^{-2/3}\gamma_0^{-5/6}. \quad (16)$$

As we have mentioned above, the average luminosity of the BB component is $L_{\rm iso,BB} \simeq 1.1\times 10^{46}\ {\rm erg\ s^{-1}}$, which is of the order of $L_{\rm ph}$ estimated here.

Therefore, based on the observed temperature and luminosity, the delay between the GW signal and the soft BB component can be accounted for as the time it takes the fireball photosphere to radiate. The primary challenge of this interpretation is in explaining the nature of the hard non-thermal emission preceding the BB component. If both components are the result of the same expanding fireball, the photospheric emission is expected to occur earlier than or at the same time as the non-thermal emission. This requirement can be reconciled with the GBM data if the thermal component was subdominant and indistinguishable during the initial hard non-thermal pulse.

Alternatively, energy dissipation below the photospheric radius could also provide an explanation for the timing of the two pulses (Rees & Mészáros 2005). This could be achieved through a range of possible scenarios. Energy dissipation could occur through inelastic collisions between the decoupled neutron and proton populations within the jet (Beloborodov 2010), for example, or through magnetic reconnection processes (Giannios 2006). The emitted radiation would exhibit a modified blackbody spectrum and be released at the photospheric radius, resulting in similar arrival times to the thermal emission. A non-thermal pulse could also arise from a forward shock if the deceleration radius were located below the photosphere. Such a scenario would be possible if the density of the external medium were sufficiently in excess of the interstellar medium, which is a distinct possibility for such environments (Goriely et al. 2011; Bauswein et al. 2013; Hotokezaka et al. 2013).

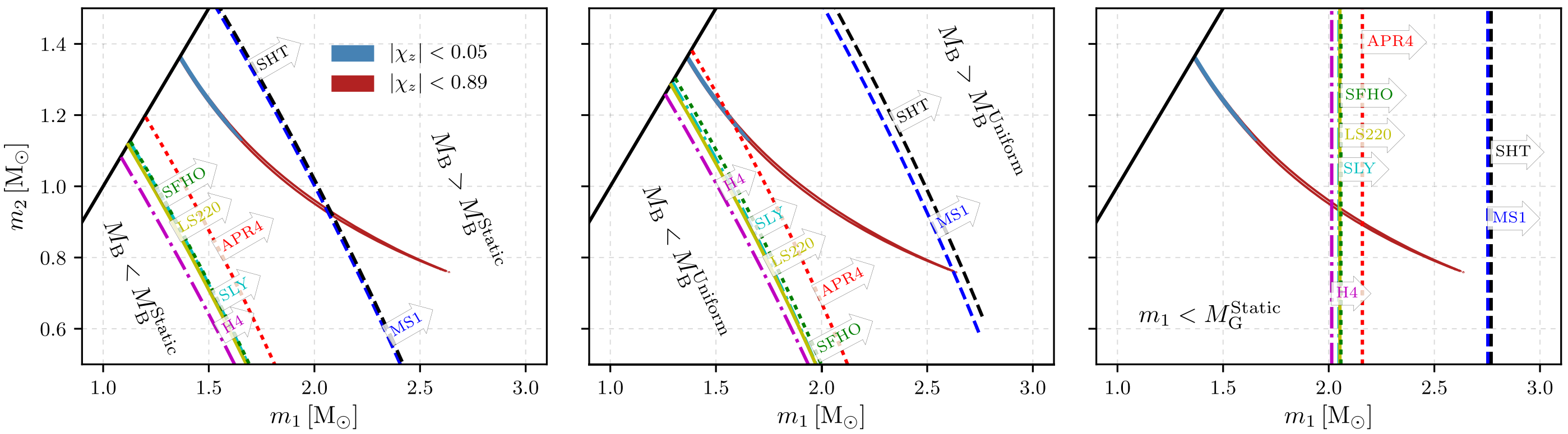

**Figure 3.** Critical mass boundaries for different EOSs in comparison with the 90% credible region of the gravitational masses inferred from GW170817 (prior limits on the spin magnitude, $|\chi_z|$, given in the legend). The slanted curves in the left panel and middle panel correspond to the maximum baryonic mass allowed for a single non-rotating NS (left) and for a uniformly rotating NS (middle). Arrows indicate for each EOS the region in the parameter space where the total initial baryonic mass exceeds the maximum mass for a single non-rotating or uniformly rotating NS, respectively. The right panel illustrates EOS-dependent cuts on the gravitational mass $m_1$ of the heavier star, with arrows indicating regions in which $m_1$ exceeds the maximum possible gravitational mass $M_G^{\rm Static}$ for non-rotating NSs. In all three panels the black solid line marks the $m_1 = m_2$ boundary, and we work in the $m_1 > m_2$ convention.

The thermal component could also be the result of "cocoon" emission from shocked material surrounding the relativistic jet (Lazzati et al. 2017), which is expected to be softer and fainter than the non-thermal prompt emission (Ramirez-Ruiz et al. 2002; Pe'er et al. 2006). To examine this scenario, we utilize the relation between the radius of the shock breakout, duration and observed temperature proposed by Nakar & Sari (2012):

$$R = 1.4 \times 10^9 \left(\frac{t}{1\ \rm s}\right)\left(\frac{T}{10\ \rm keV}\right)^2 \rm cm. \qquad (17)$$

Using the distance measurement, the blackbody spectral fit implies a radius of $R_{\rm BB} = 3 \times 10^8$ cm that we can use as a proxy for the cocoon radius. This is within a factor of 4 of the relation, in spite of the fact that it applies to spherical geometry that is not guaranteed here, and the $R_{\rm BB}$ radius derived from the fit assumes thermal equilibrium. We thus consider this as evidence supporting the cocoon scenario.

Finally, the delay between the GW and EM signals may also be due in part to the time it takes for the relativistic jet to break out of the sub-relativistic dense ejecta surrounding the merger (Nagakura et al. 2014; Moharana & Piran 2017). We estimate that the breakout time for typical dynamical ejecta mass values of $\sim$0.1 $M_\odot$ in such a merger (Hotokezaka et al. 2013) could not account for the entire observed delay. Lowering the reference isotropic kinetic luminosity of $L_{k,\rm iso} = 10^{51}$ erg s$^{-1}$ assumed by Moharana & Piran (2017) could be one way to account for a larger delay. However, a luminosity below the one assumed in the breakout scenario substantially increases the likelihood of a "choked" jet that fails to break out of the surrounding medium (Aloy et al. 2005).

### *5.2. Neutron Star EOS Constraints*

The observation of an SGRB associated with the merger of two NSs can be used to derive constraints on the EOS of NS matter (see theoretical studies by Belczynski et al. 2008; Fryer et al. 2015; Lawrence et al. 2015). To do this, we compare the measurement of the binary mass from the GW signal with two possible models of the merger remnant that powered the SGRB: (i) the merger remnant collapsed to a rotating BH with a surrounding disk that powered the SGRB (Shibata et al. 2006), or (ii) the merger formed a rapidly rotating, strongly magnetized NS (millisecond magnetar) with an accretion disk (Metzger et al. 2008).

We consider a representative sample of EOSs: SLy (Douchin & Haensel 2001), LS220 (Lattimer & Swesty 1991), SFHo (Steiner et al. 2013), H4 (Lackey et al. 2006), APR4 (Akmal et al. 1998), SHT (Shen et al. 2011), and MS1 (Müller & Serot 1996). For each EOS, we compute the maximum stable baryonic mass and gravitational mass of a non-rotating (static) NS, denoted $M_{\rm B}^{\rm Static}$ and $M_{\rm G}^{\rm Static}$, respectively, and the maximum baryonic mass of a uniformly rotating NS $M_{\rm B}^{\rm Uniform}$ (Gourgoulhon et al. 2001). The merger remnant can only collapse to a BH if its baryonic mass is larger than $M_{\rm B}^{\rm Static}$.

If we neglect rotational corrections, the baryonic masses $m_{\rm B1}$, $m_{\rm B2}$ of the initial NSs are functions of their gravitational masses $m_1$, $m_2$ only. In this approximation, a fixed total initial baryonic mass, $M_{\rm B}^{\rm Initial}$, corresponds to a curve in the ($m_1$, $m_2$) parameter space. In Figure 3 we show lines of $M_{\rm B}^{\rm Static}$ and $M_{\rm B}^{\rm Uniform}$ that bound the region of the parameter space in which the total mass of the binary is consistent with a stable non-rotating or uniformly rotating remnant, respectively. The figure also contains the 90% credible region of the gravitational masses obtained with a restricted or full spin prior (Abbott et al. 2017e). We note that the latter has a broader distribution of the component masses, such that the heavier NS can exceed $M_{\rm G}^{\rm Static}$ for various EOS, which would correspond to either a supramassive (or even hypermassive) NS, or to a light BH. The maximum gravitational masses allowed for each EOS, $M_{\rm G}^{\rm Static}$, are shown in the figure as vertical lines.

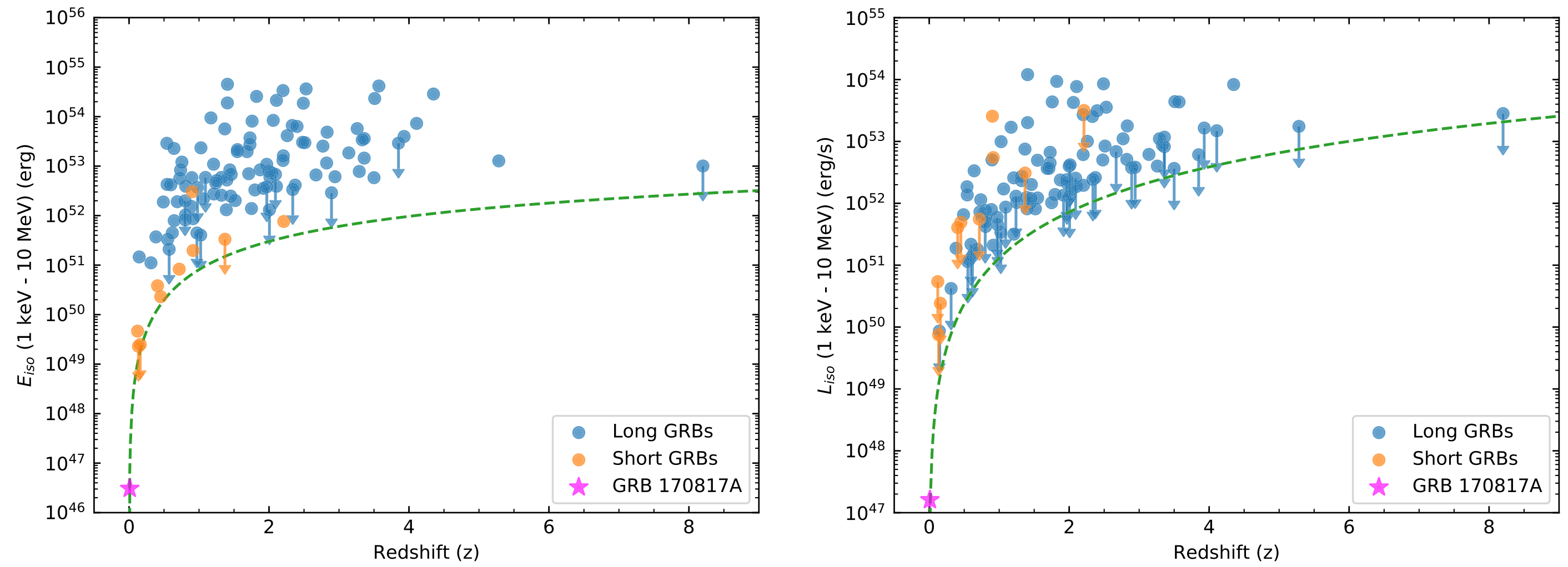

**Figure 4.** GRB 170817A is a dim outlier in the distributions of $E_{\rm iso}$ and $L_{\rm iso}$, shown as a function of redshift for all GBM-detected GRBs with measured redshifts. Redshifts are taken from GRBOX (http://www.astro.caltech.edu/grbox/grbox.php) and Fong et al. (2015). Short- and long-duration GRBs are separated by the standard $T_{90} = 2$ s threshold. For GRBs with spectra best modeled by a power law, we take this value as an upper limit, marking them with downward pointing arrows. The power law spectra lack a constraint on the curvature, which must exist, and therefore, will overestimate the total value in the extrapolated energy range. The green curve demonstrates how the (approximate) GBM detection threshold varies as a function of redshift. All quantities are calculated in the standard 1 keV–10 MeV energy band.

Since the total baryonic mass of the system can only be reduced (by mass ejection), the maximum baryonic mass of the merger remnant and accretion disc is bound by $M_{\rm B}^{\rm Initial}$. From Figure 3, we can see that for the measured NS gravitational masses with the low-spin prior, the MS1 and SHT EOS could not form a BH since $M_{\rm B}^{\rm Initial} < M_{\rm B}^{\rm Static}$. Assuming that the magnitude of the spins is small, the MS1 and SHT EOS are incompatible with BH formation. If the dimensionless spins of the NSs are allowed to be larger than 0.05, BH formation is only disfavored: we find that a fraction 83% (MS1) and 84% (SHT) of the posterior distribution satisfies $M_{\rm B}^{\rm Initial} < M_{\rm B}^{\rm Static}$. For both spin priors, we find that the H4, LS220, SFHo, and SLy EOS result in $M_{\rm B}^{\rm Initial} > M_{\rm B}^{\rm Uniform}$. Even when assuming a large ejecta mass of 0.1 $M_\odot$, the remaining mass cannot form a uniformly rotating NS. For those EOS, the merger either results in prompt BH formation or in a short-lived remnant, with a lifetime determined by the dissipation of differential rotation and/or disk accretion.

To be compatible with scenario (ii), the lifetime of the merger remnant would have to be sufficiently long to power the GRB. We note that prompt BH formation is a dynamic process accessible only to numerical relativity simulations. Although there are parameter studies (Hotokezaka et al. 2011; Bauswein et al. 2013), they only consider equal mass binaries. Considering also the error margins of those studies, we currently cannot exclude prompt collapse for the H4, LS220, SFHo, and SLy EOS. Finally, we note that for the APR4 EOS only the possibility of a stable remnant can be ruled out. More generally, only EOSs with $M_{\rm B}^{\rm Static} < 3.2\,M_\odot$ are consistent with scenario (i) when assuming the low-spin prior, or with $M_{\rm B}^{\rm Static} < 3.7\,M_\odot$ for the wider spin prior. These bounds were derived from the 90% credible intervals of the $M_{\rm B}^{\rm Initial}$ posteriors (and these, in turn, are determined for each EOS in order to account for binding energy variations). These upper limits are compatible with and complement the lower bounds on $M_{\rm G}^{\rm Static}$ from the observation of the most massive known pulsar, which has a mass of $(2.01 \pm 0.04)\,M_\odot$ (Antoniadis et al. 2013). In Section 6.5 we will discuss some model-dependent implications of the lack of precursor and temporally extended gamma-ray emission from GRB 170817A on the progenitor NSs.

## 6. Gamma-ray Energetics of GRB 170817A and their Implications

Using the measured gamma-ray energy spectrum and the distance to the host galaxy identified by the associated optical transient, we compare the energetics of GRB 170817A to those of other SGRBs at known redshifts. Finding GRB 170817A to be subluminous, we discuss whether this dimness is an expected observational bias for joint GW–GRB detections, what insight it provides regarding the geometry of the gamma-ray emitting region, what we can learn about the population of SGRBs, update our joint detection estimates, and set limits on gamma-ray precursor and extended emission.

### *6.1. Isotropic Luminosity and Energetics of GRB 170817A*

Using the "standard" spectral information from Goldstein et al. (2017) and the distance to the host galaxy NGC 4993 $(42.9 \pm 3.2)$ Mpc, we calculate the energetics of GRB 170817A using the standard formalisms (Bloom et al. 2001; Schaefer 2007). GRBs are believed to be relativistically beamed and their emission collimated (Rhoads 1999). Isotropic energetics are upper bounds on the true total energetics assuming the GRB is observed within the beaming angle of the brightest part of the jet. We estimate that the isotropic energy release in gamma-rays $E_{\rm iso} = (3.1 \pm 0.7) \times 10^{46}$ erg, and the isotropic peak luminosity, $L_{\rm iso} = (1.6 \pm 0.6) \times 10^{47}\,{\rm erg\,s^{-1}}$, in the 1 keV–10 MeV energy band. These energetics are from the source interval—i.e., the selected time range the analysis is run over—determined in the standard manner for GBM spectral catalog results, allowing us to compare GRB 170817A to other GRBs throughout this section. The uncertainties on the inferred isotropic energetics values here include the uncertainty on the distance to the host galaxy. As a cross check, the isotropic luminosity is also estimated using a Bayesian approach proposed by Fan (2017). Assuming a flat prior on isotropic luminosity, we obtain $L_{\rm iso} = 1.2^{+0.7}_{-0.6} \times 10^{47}$ erg s$^{-1}$, which is consistent with the standard GBM approach. This Bayesian approach can be used to combine future joint GW-GRB observations to provide a redshift-independent estimate of the GRB luminosity function.

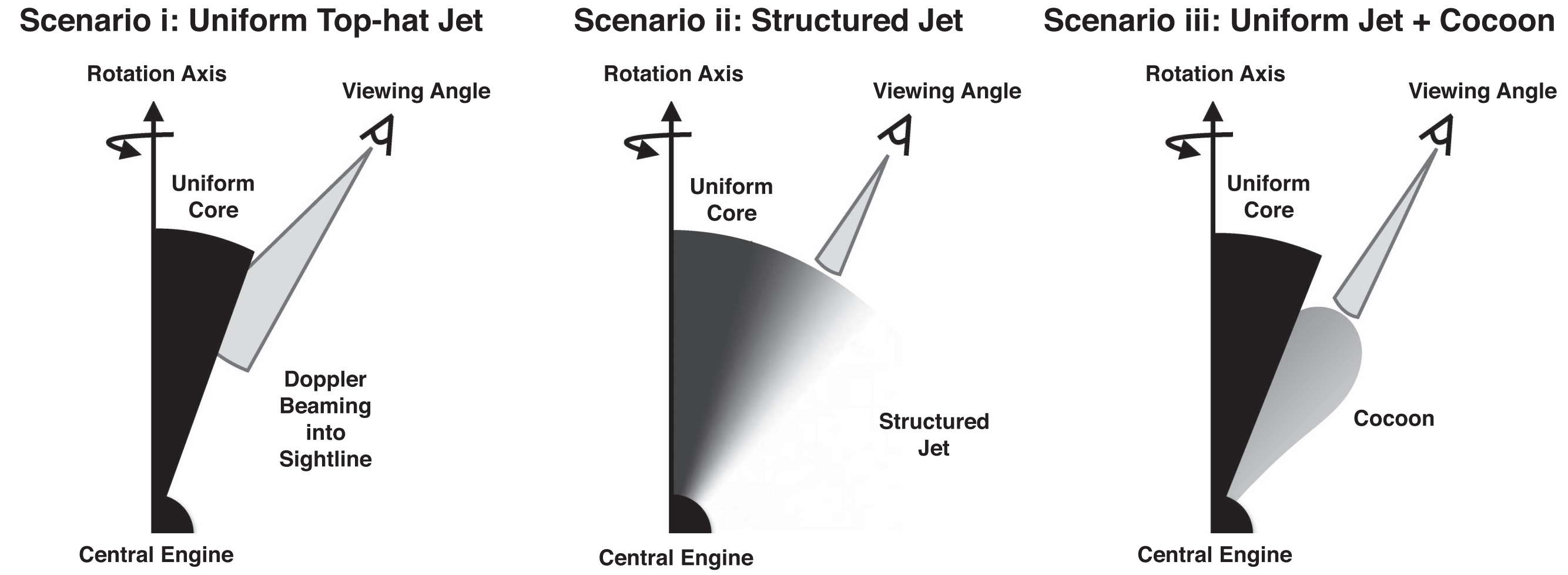

**Figure 5.** Three potential jet viewing geometries and jet profiles that could explain the observed properties of GRB 170817A, as described by scenarios (i)–(iii) in Section 6.2.

The two apparent components of GRB 170817A are sufficiently different that using an average spectrum to estimate the fluence may produce an inaccurate total luminosity. Therefore, we also estimate $E_{\rm iso}$ using the "detailed" fits described in Goldstein et al. (2017). Separating the hard peak best fit by a Comptonized function (a power law with an exponential cutoff) and the softer tail best fit by a BB spectrum, we estimate $E_{\rm iso,comp} = (4.0 \pm 1.0) \times 10^{46}$ erg, and $E_{\rm iso,BB} = (1.3 \pm 0.3) \times 10^{46}$ erg, for a total of $E_{\rm iso} = (5.3 \pm 1.0) \times 10^{46}$ erg.

Compared to the distribution of GBM detected GRBs with measured redshift shown in Figure 4, GRB 170817A is 2 orders of magnitude closer and 2 to 6 orders of magnitude less energetic than other SGRBs. In particular, GRB 150101B was previously the weakest SGRB with a firm redshift association ($z = 0.134$; Fong et al. 2016), and its energetics (as measured by GBM) $E_{\rm iso} = 2.3 \times 10^{49}$ erg, and $L_{\rm iso} = 7.5 \times 10^{49}$ erg s$^{-1}$ are 2–3 orders of magnitude higher. As this was the previous dimmest burst, the minimum luminosity cut of $5 \times 10^{49}$ erg s$^{-1}$ used in Wanderman & Piran (2015) to fit a rate and an $L_{\rm iso}$ distribution to existing observations appeared reasonable; however, with GRB 170817A, the lower bound on the isotropic energetics distributions needs to be revised, as discussed in Section 6.4.

### *6.2. Implications of the Dimness on the Central Engine*

The broad observed brightness distribution likely arises from a mixture of an intrinsic brightness distribution and geometric effects, which include the inclination angle of the system to Earth, the structure and width of the collimated jet itself, and the relativistic beaming angle $\theta_b$. We consider several possibilities to explain why GRB 170817A is extremely dim (Figure 5): (i) we viewed it from beyond the half-jet opening angle $\theta_j$ for a standard top-hat model, (ii) the structure of the jet is more complicated than a simple top-hat model, (iii) the observed emission for GRB 170817A originates from a different mechanism than for most SGRBs, or (iv) it is due solely to the intrinsic luminosity distribution and not the geometry of the system.

*Scenario (i).* Uniform top-hat jets (constant emissivity and Lorentz factor, $\Gamma$, within the jet aperture) with a sharp edge have been widely used to explain GRB properties, including jet breaks (Rhoads 1999). The top-hat jet is the simplest possible model for calculating off-axis parameters as it captures the basic physics of the system, but it is unable to account for smooth profiles in the Lorentz factor and the emissivity. Here the observed energetics are significantly lower than they would be if we were within $\theta_j$.

In the top-hat scenario, off-axis values of physical quantities can be related to the on-axis values through the angle dependence of the relativistic Doppler factor:

$$\delta_{\rm D}(\theta) = [\Gamma(1 - \beta\cos\theta)]^{-1} \approx 2\Gamma/(1 + \theta^2\Gamma^2), \quad (18)$$

where $\theta$ is the angle between the velocity vector $v$ and the line of sight, and $\beta = v/c$. The relation for duration and peak energy is linear with $\delta_{\rm D}$ (see, e.g., Granot et al. 2002):

$$\frac{T_{90}({\rm off-axis})}{T_{90}({\rm on-axis})} = \frac{E_{\rm p}({\rm on-axis})}{E_{\rm p}({\rm off-axis})} = \frac{\delta_{\rm D}(0)}{\delta_{\rm D}(\theta_{\rm j} - \zeta)} = \frac{1 - \beta\cos(\theta_{\rm j} - \zeta)}{1 - \beta} \stackrel{\Delta}{=} b \approx 1 + \Gamma^2(\zeta - \theta_{\rm j})^2, \quad (19)$$

whereas $E_{\gamma,\rm iso}$(off-axis) scales approximately $\propto b^{-2}$ for a viewing angle $\zeta$ between $\theta_{\rm j}$ and $2\theta_{\rm j}$. The duration in the on-axis scenario may be longer than inferred from the above equation, as the variable gamma-ray flux can be discerned above detector noise for a longer fraction of the total activity compared to emission viewed off-axis.

We use the observed quantities for GRB 170817A, $E_p \approx 200$ keV, $E_{\gamma,\rm iso} = 5.3 \times 10^{46}$ erg, and $T_{90} \approx 2$ s, as values observed off-axis. If we assume that the on-axis values for GRB 170817A are consistent with typical values observed for SGRBs, we obtain $E_p = 6(b/30)$ MeV, $E_{\gamma,\rm iso} = 5 \times 10^{49}(b/30)^2$ erg, and $T_{90} = 7 \times 10^{-2}(b/30)^{-1}$ s. In particular using a fiducial range on $E_{\gamma,\rm iso}({\rm on-axis})$ corresponding to the two orders of magnitude spread shown in Figure 4 we obtain $b \approx \Gamma^2(\zeta - \theta_j)^2 \approx 30$ within a factor 3, which is a constraint on the values of $\Gamma$, $\zeta$ and $\theta_j$.

If we assume a viewing angle of $\zeta = 30°$ and $\Gamma = 300$ the uncertainty on $b$ yields $\zeta - \theta_j \simeq 1 \pm 0.5$ deg, a solid angle covering only 1% of a full sphere. Hence this configuration would require a fine tuning of the line of sight. However, if we assume $\Gamma = 30$ then the uncertainty on $b$ yields $\zeta - \theta_j \simeq 10 \pm 4$ deg, a solid angle that covers 10% of a full sphere, which is plausible without too much fine tuning. This argument only weakly depends on the particular value $\zeta$, and illustrates that for large $\Gamma$ a top-hat jet scenario is disfavored due to the sharp emission fall-off at the edges.

*Scenario (ii).* A more complex geometry involves a structured jet (Rossi et al. 2002, or Granot 2007 and references therein) which provides a wider range of angles from which the observer could still detect emission, and therefore does not require a fine-tuned viewing angle. Structured jet emission profiles include a uniform ultra-relativistic core surrounded by a power-law decaying wing where the energy and Lorentz factor depend on the distance from the jet axis (Pescalli et al. 2015), a Gaussian with a smooth edge and falloff outside the core (Zhang & Mészáros 2002; Kumar & Granot 2003), and a two-component jet with an ultra-relativistic narrow core and slightly slower outer jet (Frail et al. 2000; Berger et al. 2003; Racusin et al. 2008; Filgas et al. 2011), among other possibilities.

Structured jets can naturally explain the broad observed energetics distribution. Because SGRBs involve relativistic velocities, radiation is strongly beamed into angle $\theta_b$. If the observed brightness depends on viewing geometry, i.e., is not uniform across the angle $\theta_j$, then the part of the beam that we observe may be off-axis to the brightest part of the jet but we may still be within $\theta_b$ of some dimmer part of the emitting region, though in this case we would expect the $\Gamma$ factor to vary as well.

*Scenario (iii).* Given the closeness of this burst it is possible that the observed emission is due to a different mechanism from other SGRBs, one that is intrinsically dim and thus undetectable at usual SGRB distances. We believe this explanation to be unlikely as the main emission episode of GRB 170817A is a typical SGRB (as measured by the observed gamma-ray properties). It is possible that the soft tail emission arises from a distinct mechanism. One explanation is "cocoon" emission from the relativistic jet shocking its surrounding non-relativistic material (Lazzati et al. 2017). We showed that "cocoon" emission could explain the thermal tail in Section 5.1. A possible full model for GRB 170817A is off-axis emission from a top-hat jet providing the main emission episode, with "cocoon" emission arising from the jet's interaction with the surrounding torus that powers the main jet. The softer emission is near the detection limits of GBM and would not be detected to much greater distances, suggesting it may be a common property of SGRBs that is otherwise missed.

*Scenario (iv).* If GRB 170817A is viewed within both the collimated jet and the beaming angle, and the emission is constant across the traditional top-hat jet, then GRB 170817A is intrinsically much dimmer by orders of magnitude compared to other observed GRBs. This would mean that top-hat jets have an intrinsic distribution covering 6 orders of magnitude, which is difficult to envision given the limited mass ranges in the merger of two NSs (although see Metzger & Berger (2012) and references therein). A broader intrinsic luminosity distribution might be accommodated if we assume that at least some SGRBs arise from the merging of an NS with a BH. It is possible, for example, that the brightest events may arise from NS–BH mergers with optimal mass ratio and spin parameters. Another possibility is that this broad luminosity range could arise from other properties of the system, such as the magnetic field strength of the progenitors or the intrinsic jet-opening angle distribution.

Observations of GW170817/GRB 170817A at other wavelengths (which are not explored in this *Letter*) will be necessary for a full understanding of this event. For example, evidence for X-ray emission that only arises at late times may provide evidence for this event occurring off-axis (see, e.g., Mészáros et al. 1998; Granot et al. 2002; Yamazaki et al. 2002). However, future joint detections of GW-GRB events can also provide a fuller understanding of the intrinsic energetics distributions and the effect geometry has on our observed brightness. Here the inclination constraint is not particularly informative as the inclination angle constraint, $\zeta \leqslant 36°$, is comparable to the highest lower limit for a half-jet opening angle, $\theta_j > 25$ deg (Fong et al. 2015). If this is truly off-axis from a top-hat jet then it is unlikely to be a common occurrence. Only joint GW–EM detections will reveal if the intrinsic brightness varies according to the type of progenitor.

The updated expected joint detection rates in Section 6.4 suggest inferences on populations of joint detections may be possible sooner than previously thought.

### 6.3. Observational Bias Against Low-luminosity GRBs

The fact that GRB 170817A is orders of magnitude dimmer than the population of SGRBs with known redshifts raises the questions: (i) is it unexpectedly dim, and (ii) is there a population of SGRBs with comparable luminosities (and distances) that we are not detecting? We explore here whether the gap in luminosity compared to more luminous SGRBs is a result of the instrumental sensitivity for the detection of either the prompt or the afterglow emission of SGRBs, or whether our problem lies in the association of SGRBs to their host galaxies and thus redshift.

Burns et al. (2016) examine the observed relationship between redshift and gamma-ray fluence for SGRBs with known redshift and find no strong correlation. SGRBs that appear extremely bright are likely to be nearby because their inferred luminosities would otherwise be unrealistic, but SGRBs near the detection threshold of (current) GRB detectors are as likely to be nearby as far away. The intrinsically dim part of the SGRB luminosity distribution is detectable only at short distances.

GRB 170817A is our only clear case of a subluminous SGRB with known distance, so we investigate the maximum distance at which it could have triggered GBM. Assuming the event occurred at the same time and viewing geometry with respect to *Fermi*, with comparable detector background rates, we find that GRB 170817A could have been ~30% dimmer before falling below the on-board triggering threshold (Goldstein et al. 2017), corresponding to a maximum detection distance of about 50 Mpc. An approximate measure of the detectability distance given optimal detection conditions (e.g., low background, good geometry) suggests the maximum distance we could have detected this burst is about 80 Mpc—closer than any other SGRB with a firmly determined redshift. While the GW horizon has been considered the limiting factor for joint detections with EM signals, this joint detection shows that we now must also account for an SGRB detection horizon given the sensitivity of the current gamma-ray observatories.

In addition to limited gamma-ray detector sensitivity, determining the redshift from EM observations alone is more difficult for SGRBs than for long GRBs. The fraction of SGRBs with detected X-ray afterglows for *Swift* BAT detected bursts is ∼75% (Fong et al. 2015), compared to over 90% for long GRBs.[185] It is possible that SGRBs with subluminous prompt gamma-ray emission also have correspondingly weaker X-ray afterglows, and these could account for a large fraction of the quarter of SGRBs without X-ray detections. Even when the X-ray afterglows are detected, they are fainter and thus fade below detectability threshold faster than the afterglows of long GRBs, making direct measurement of the redshift from the afterglow exceedingly rare (Fong et al. 2015). For SGRBs, the redshift is instead usually determined from the host galaxy. This requires first that the afterglow be tied to a particular host galaxy, which can be difficult because the SGRB progenitors sometimes lie outside their putative hosts, owing to the natal kicks induced by the supernovae that produced the compact objects in the progenitor system (Wong et al. 2010). A well-localized (∼few arcseconds) SGRB afterglow is associated with a galaxy within a small angular distance on the sky, using probabilistic arguments about chance alignment, and then the redshift of the host galaxy is measured.

Appendix B lists all SGRBs with possible redshifts. Most of the list was compiled by combining three relatively complete and recent literature samples (Fong et al. 2015; Lien et al. 2016; Siellez et al. 2016). Nearly all of these were detected by *Swift* BAT. It has been suggested that the BAT SGRB distribution is contaminated by the short tail of the long GRB distribution (Bromberg et al. 2013). Burns et al. (2016) find that the BAT sample is not significantly more contaminated than the GBM sample and the redshift distribution based on *Swift* BAT SGRBs is therefore a valid proxy for the redshift distribution of GBM SGRBs in the following discussion.

Berger (2010) discuss "hostless" SGRBs, which are well-localized SGRBs that have no obvious associated host galaxy despite deep observational limits. They suggested the hosts could be nearby galaxies at larger angular offset to the afterglow than others farther away, but also put forward the possibility of more distant, undetected hosts. Tunnicliffe et al. (2014) show that hostless SGRBs have an excess of nearby galaxies within a few arcminutes, relative to long GRBs or random positions, suggesting that at least some of these hostless SGRBs have nearby hosts. Therefore, the traditional assignment of probability of an SGRB to a host galaxy based solely on angular offset from a well-localized afterglow may exclude real associations with larger offsets, which are more likely to be measured for nearby events.

Tunnicliffe et al. (2014) also includes the closest potential host for an SGRB prior to GRB 170817A, with 81 Mpc for GRB 111020A. If real, this association implies an extremely low $E_{iso}$ of $\sim 10^{46}$ erg, similar to GRB 170817A. In light of the secure association of GRB 170817A with GW170817, a subenergetic $E_{iso}$ may no longer be a reason to doubt subluminous nearby SGRB host associations, and may suggest a reconsideration of very nearby host galaxies with large projected angular offsets for hostless SGRBs. We include these putative host associations in Table 2. Also included are SGRBs with extended emission and bursts that have durations exceeding the standard $T_{90}$ cut but are believed to be short based on other evidence such as spectral hardness. Asterisks indicate bursts, where the *Swift* BAT $T_{90} < 2$ s, that have localizations of a few arcseconds or better (as larger localizations increase the chance of false associations due to chance alignment), and for which the angular offset of the afterglow from the host fulfills standard association criteria. For further analysis this restricted sample is our "gold sample," and the full sample the "total sample."

One outstanding question is why we have not detected other SGRBs as close as GRB 170817A. We have established that bursts as dim as GRB 170817A will not be detected by current gamma-ray instruments if they lie much farther away than GRB 170817A. This raises questions about GRBs with luminosities between GRB 170817A and the rest of the GRBs with known luminosities. Some of these are surely being detected, albeit with unassigned redshifts and thus luminosities. While there are only ∼40 SGRBs with possible redshifts, several hundreds have been detected without an assigned redshift. Nearby, subluminous SGRBs surely lie among them. There is a lower priority for following-up weak SGRBs, so if nearby events are systematically detected as weak bursts they may not have the required follow-up observations at lower wavelengths to determine the distance to the burst. It could also be that these weak bursts also have lower brightness at lower wavelengths, making them harder to detect even with follow-up observations. Lastly, we could be detecting these bursts in gamma-rays and X-rays, but failing to properly associate them with their hosts as discussed above.

GRB 170817A is unique in that its distance was first measured by GWs, which are currently detectable out to limited distances (roughly 100 Mpc) compared to other SGRBs with known redshifts (see Table 2). This is analogous to the first association of long GRBs with supernovae. Long GRBs have redshifts systematically higher than SGRBs (Coward et al. 2013). The long burst GRB 980425 is the closest GRB to date with a measured distance (and the only GRB of any class closer than GRB 170817A), and it was the first long GRB associated with a supernovae. GRB 980425 was 4 orders of magnitude less energetic than other GRBs detected at that time (Galama et al. 1998). Because supernovae are less luminous than long GRBs, the long GRBs that are associated with supernovae are systematically closer than the average population. Because of the Malmquist bias, a bias toward detecting intrinsically bright objects (Malmquist 1922), we only see dim long GRBs when they are nearby. This explains the subluminous nature of GRB 980425, and this observational peculiarity has been confirmed by other subluminous long GRB-SN detections, including GRB 031203/SN2003lw (Malesani et al. 2004), GRB 060218/SN2006aj (Modjaz et al. 2006), and GRB 100316D/SN2010bh (Cano et al. 2011), quantified as a population in Howell & Coward (2012). The history of GRB 980425, the other nearby subluminous long GRBs associated with supernova, and the lack of correlation between SGRB gamma-ray fluence and redshift noted by Burns et al. (2016) motivates the further development of subthreshold searches for counterparts to GW events and for subthreshold SGRBs in general. While GRB 170817A occurred nearby, and its favorable geometry to *Fermi* resulted in an on-board trigger, we anticipate that these untriggered searches of GBM and other gamma-ray data will uncover future counterparts to this GW-selected SGRB population.

[185] https://swift.gsfc.nasa.gov/archive/grb_table/

Suggestions of nearby subluminous SGRB populations existed prior to this discovery (Tanvir et al. 2005; Siellez et al. 2016). Tanvir et al. (2005) find a statistically significant correlation between a large sample of coarsely localized SGRBs detected by the Burst And Transient Source Experiment (BATSE) and a sample of nearby galaxies. Without associating individual SGRBs with a potential host, they conclude that ∼10% of the SGRB sample could be part of a nearby subluminous population. Siellez et al. (2016) infer the presence of a nearby subluminous SGRB population through a study of SGRBs with known redshift in the context of BNS and NS–BH population evolution. They find an excess of actual nearby low-luminosity SGRBs using the results from their simulations, covering a broad range of assumed lifetimes for the binary system prior to merger.

Giant flares from the highly magnetized NSs known as magnetars can be detected outside our galaxy, with the sole extragalactic example tied to its host coming from SGR 0525-66, in the Large Magellanic Cloud (Evans et al. 1980). A giant flare from the galactic magnetar SGR 1806-20 showed a gamma-ray spectrum measured by the Konus–Wind instrument that was well-fit by a blackbody with temperature ∼175 keV (Hurley et al. 2005), harder than a regular magnetar burst. This hard spectrum led to the idea that giant flares from magnetars in nearby galaxies might be a sub-population hiding among the general SGRB population (Hurley et al. 2005). Tanvir et al. (2005) found a stronger correlation of BATSE SGRBs with early-type than late-type galaxies, which is not expected if nearby SGRBs arise from giant magnetar flares in nearby galaxies, but is consistent with a BNS origin. GRB 170817A is clearly associated with a BNS merger, but even without the connection to GW170817, the spectrum of GRB 170817A in the GBM data strongly disfavored the BB fit expected for a giant magnetar flare. Another possible signature of a giant magnetar flare is the ringing in its tail at the NS period of a few seconds, which could be detected by GBM or by SPI-ACS for flares outside our galaxy providing it was close enough. A search for periodic or quasi-periodic emission in the GBM data for GRB 170817A (Goldstein et al. 2017) found no periodic modulation, providing another discriminant between SGRBs and nearby extragalactic giant magnetar flares that might be masquerading as SGRBs.

### *6.4. Predicted Detection Rates*

The intrinsic specific volumetric SGRB rate is often quoted to be around $10\,\mathrm{Gpc}^{-3}\,\mathrm{yr}^{-1}$ (see, e.g., Guetta & Piran 2006; Coward et al. 2012; Fong et al. 2015; with the value originating from Nakar et al. 2006, who noted that the true rate could be much higher). However, unlike GW signals, SGRBs do not have a clear relationship between the observed distance and brightness. As discussed in the previous section, this can be due to intrinsic variations in SGRB luminosities, as well as structure in the jet. In this Section, we investigate the former scenario presenting the implications of GW170817/GRB 170817A for future GW and SGRB observations in terms of a simple standard model for the SGRB luminosity distribution. Similar interpretations for other, perhaps more elaborate, models are straightforward.

We model the SGRB luminosity function as a broken power law, with a logarithmic distribution[186]

$$\phi_o(L_{\mathrm{iso}}) = \begin{cases} \left(\dfrac{L_{\mathrm{iso}}}{L_\star}\right)^{-\alpha_L} & L_{\mathrm{iso}} < L_\star \\ \left(\dfrac{L_{\mathrm{iso}}}{L_\star}\right)^{-\beta_L} & L_{\mathrm{iso}} > L_\star \end{cases}, \tag{20}$$

where $L_{\mathrm{iso}}$ is the peak isotropic luminosity (in the source frame) between 1 keV and 10 MeV, and $\alpha_L$ and $\beta_L$ give the power law decay below and above the break at $L_\star$.[187] Here, we follow Wanderman & Piran (2015) in using $L_\star \simeq 2 \times 10^{52}$ erg s$^{-1}$, $\alpha_L \simeq 1$, and $\beta_L \simeq 2$. The other important parameter is the minimum SGRB luminosity, which determines the lower cutoff of the luminosity distribution. This is poorly constrained as only nearby low luminosity SGRBs are observable. In Wanderman & Piran (2015) the minimum luminosity is taken to be $L_{\mathrm{min}} = 5 \times 10^{49}$ erg s$^{-1}$, while other studies use values ranging from $1 \times 10^{49}$ erg s$^{-1}$ to few $\times\, 10^{50}$ erg s$^{-1}$ (Regimbau et al. 2015). We assume a threshold value for detectability in GBM of 2 photons cm$^{-2}$ s$^{-1}$ for the 64 ms peak photon flux in the 50–300 keV band, which is higher than the minimum detectability value to account for the sky-dependent sensitivity of GBM. Furthermore, we model the SGRB spectrum using the Band function with parameters taken from Wanderman & Piran (2015) (namely, $E_{\mathrm{peak}} = 800$ keV, $\alpha_{\mathrm{Band}} = -0.5$, and $\beta_{\mathrm{Band}} = -2.25$). This spectrum is significantly harder than the one observed for GRB 170817A. The cumulative observed rate predicted for GBM by this base model is shown as a function of redshift in Figure 6 by the purple solid curve.

As discussed in Section 6.1, the inferred $L_{\mathrm{iso}}$ is $(1.6 \pm 0.6) \times 10^{47}$ erg s$^{-1}$, which is significantly lower than any previously detected SGRB, and thus is in tension with this model. In particular, we must extend the lower limit of the luminosity down by a factor of at least 500. At present, there is rather little information available about the low luminosity distribution due to the observational biases discussed in Section 6.3 and, consequently, there is a significant degeneracy between the minimum SGRB luminosity and the rate (Wanderman & Piran 2015). Let us consider the most straightforward extension of the above model and set $L_{\mathrm{min}} = 1 \times 10^{47}$ erg s$^{-1}$ while maintaining $\alpha_L = 1$. In order to retain the same prediction for high-luminosity SGRBs, this requires a 500-fold increase in the number of SGRBs, with the majority emitting at low luminosity. The cumulative observed rate predicted for GBM by this simple extension is shown as a function of redshift in Figure 6 by the red solid curve, and is comparable to the measured BNS merger rate shown in black. This simple extension would imply SGRBs are not beamed and that essentially all BNS mergers are accompanied by at least a subluminous SGRB.

Therefore, to reduce this tension and explore other possible extensions, we introduce an additional power law break below

[186] To get the linear distribution of luminosities, both $\alpha_L$ and $\beta_L$ must be increased by 1.

[187] Other studies use a smaller energy band when defining the luminosity, and this has an impact on the value of $L_\star$, although not on the slopes of the power law components.

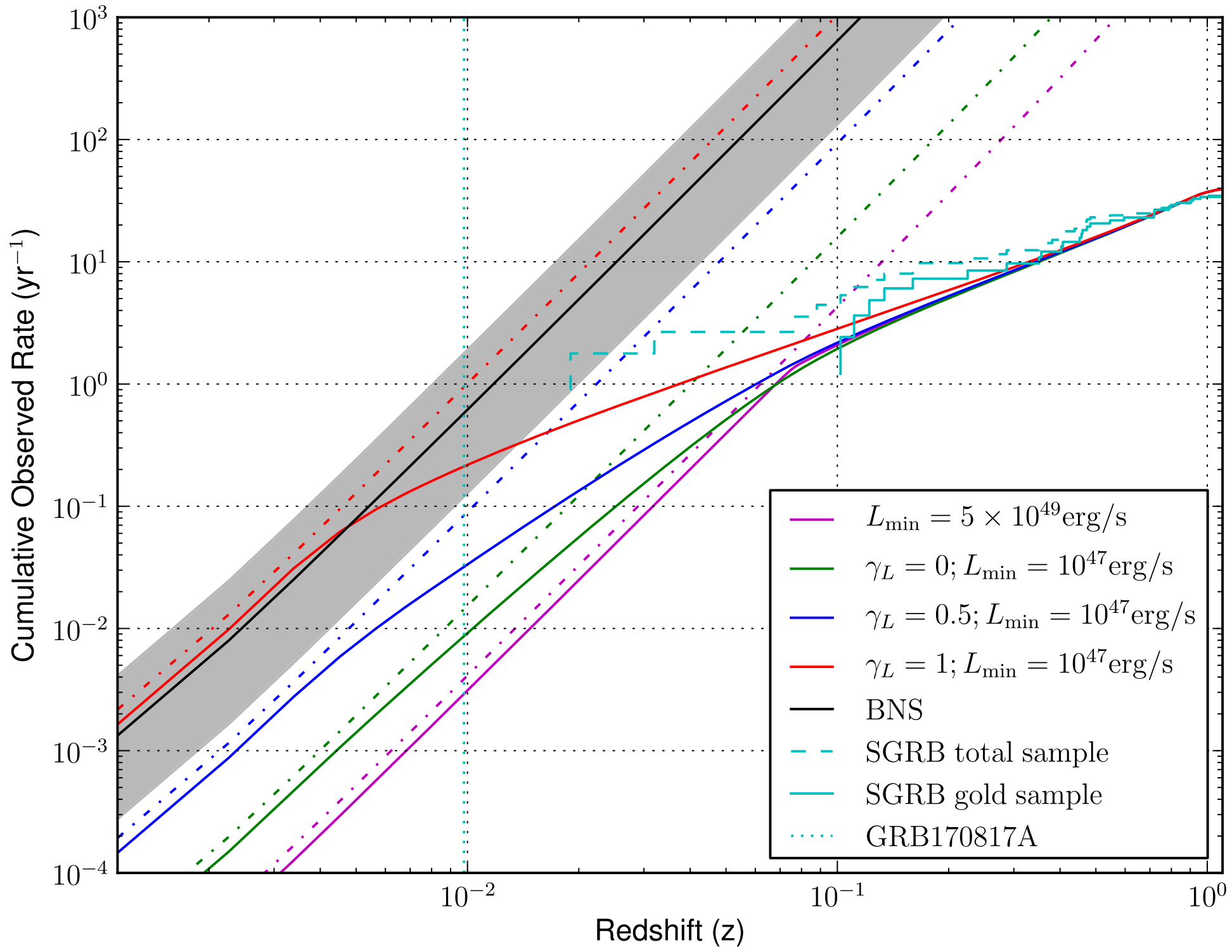

**Figure 6.** Predicted detection rates per year as a function of redshift. The red, blue, and green solid lines refer to the GBM observed SGRB rate assuming a minimum luminosity $L_{\min}$ of $1 \times 10^{47}$ erg s$^{-1}$, and $\alpha_L = 1$, $\beta_L = 2$ and $\gamma_L = \{1, 0.5, 0\}$ in Equation (21), respectively. The purple solid line refers to the base model with $L_{\min}$ of $5 \times 10^{49}$ erg s$^{-1}$. The four curves are normalized by imposing 40 triggered SGRB per year. As $\gamma_L$ increases, the observed rate is no longer volumetric at lower and lower redshifts, because a fraction of SGRBs becomes too dim to be detected. For reference, the red, blue and green dot-dashed curves show the local SGRB occurrence rate for $L_{\min} = 1 \times 10^{47}$ erg s$^{-1}$ and $\gamma_L = \{1, 0.5, 0\}$, respectively. The black line and gray band show the BNS merger rate $1540^{+3200}_{-1220}$ Gpc$^{-3}$yr$^{-1}$ determined with the detection of GW170817 (Abbott et al. 2017e). For comparison, the measured SGRBs redshift distribution from Table 2 is shown in cyan, and is broadly compatible with all of the models. The dotted vertical cyan line refers to the redshift of GRB 170817A host galaxy.

$L_{\star\star} = 5 \times 10^{49}$ erg s$^{-1}$:

$$\phi_o(L_{\rm iso}) = \begin{cases} \left(\frac{L_{\rm iso}}{L_{\star\star}}\right)^{-\gamma_L}\left(\frac{L_{\star\star}}{L_\star}\right)^{-\alpha_L} & L_{\rm iso} < L_{\star\star} \\ \left(\frac{L_{\rm iso}}{L_\star}\right)^{-\alpha_L} & L_{\star\star} < L_{\rm iso} < L_\star, \\ \left(\frac{L_{\rm iso}}{L_\star}\right)^{-\beta_L} & L_{\rm iso} > L_\star \end{cases} \quad (21)$$

and consider three values $\gamma_L = \{1, 0.5, 0\}$ for the power law index below this second break. We normalize these three cases to 40 triggered SGRBs per year for GBM, and the $\gamma_L = 1$ case corresponds to the simple extension discussed above. For reference, Figure 6 shows the local SGRB occurrence rate for $L_{\min} = 1 \times 10^{47}$ erg s$^{-1}$ and $\gamma_L = \{1, 0.5, 0\}$ indicated by the red, blue, and green dotted curves, respectively, and the BNS merger rate $1540^{+3200}_{-1220}$ Gpc$^{-3}$yr$^{-1}$ determined with the detection of GW170817 (gray band, with the mean in black; Abbott et al. 2017e). The $\gamma_L = 1$ case produces the largest number of subluminous SGRBs, and leads to a sharp departure at redshift $\sim$0.005 from volumetric detection to detection limited by the GBM sensitivity. For $\gamma_L = 0$ the transition is smoother as there is only a small number of subluminous SGRBs, and the observed rate departs gradually from the occurrence rate.

When we include the luminosity of GRB 170817A by setting $L_{\min} = 1 \times 10^{47}$ erg s$^{-1}$ and $\gamma_L = 1$, the expected detection rate at a redshift of $z \approx 0.1$ is around a factor of 2 higher than for the Wanderman & Piran (2015) model. At a redshift of $z \approx 0.01$, which is close to the observed redshift for GRB 170817A, rather than expecting to observe 1 event per 650 years with GBM, this is increased to 1 per year. The expected detection rate at a this redshift for the $\gamma_L = 0.5$ and $\gamma_L = 0$ extensions is of roughly 1 observed event per 10 and 65 years, respectively. The expectations we obtain for GBM are consistent with the distribution of SGRBs with known redshifts reported in Table 2, in Appendix B.

Using the BNS merger volumetric rate estimated from GW170817 as a new input to the detection rate calculation presented in Abbott et al. (2017a), the LIGO–Virgo detection rate is narrowed down from 0.04–100 to $\sim$1–50 BNS coalescences during the 2018–19 observing run, with the remaining uncertainty arising in part from the not-yet-known detector sensitivities during that run. At design sensitivity, the LIGO and Virgo detectors can expect to detect $\sim$6–120 BNS coalescences per year, as opposed to the previously estimated 0.1–200 BNS coalescences per year. Inclusion of any additional BNS detections in the meantime will allow this prediction to be further sharpened.

Independently, we use the GBM detection rate as a function of redshift to predict joint GW–GRB BNS detection rates (Clark et al. 2015). Both the rates and their relative uncertainties are significantly reduced, compared to the GW-only detection rate estimates above, since the majority of distant mergers will be undetectable by GBM and the GBM

SGRB detection rate is well measured. The degree to which GBM-selected SGRBs are preferentially on-axis is unclear. When estimating the joint detection rate we include both no selection and a pure on-axis selection hypothesis, the latter implying a larger detection probability by the LIGO and Virgo detectors. During the 2018–19 observing run, we expect 0.1–1.4 joint detections per year for the GW interferometer network and GBM triggered SGRBs, with the high end of the interval corresponding to the $\alpha_L = 1$, $L_{\rm min} = 1 \times 10^{47}$ erg s$^{-1}$ extension of the luminosity function. At design sensitivity, the expected joint detection rate increases to 0.3–1.7 per year.

Future joint GW-SGRB observations will provide significant new insights into low-luminosity SGRBs. In particular, both joint observations and lower limits on distances to SGRBs not observed in GWs (e.g., GRB 150906B in Abbott et al. 2017b) will constrain the rate of nearby GRBs. Future GW observations of BNS mergers will reduce the uncertainty in the rate of such events, while observation of GW signals with no SGRB counterpart will limit the SGRB beaming angles. Finally, subthreshold searches in GRB data around the time of GW events could significantly increase the number of joint observations.

### *6.5. Limits on Precursor and Extended Emission*

At gamma-ray energies, SGRBs are characterized by a prompt emission episode lasting at most ∼2 s. Observational evidence for precursor flares associated with SGRBs (Troja et al. 2010; Burns 2017; Minaev & Pozanenko 2017) and temporally extended emission (Lazzati et al. 2001; Connaughton 2002) is so far inconclusive. Given the small distance to the source, the absence of such emission from GRB 170817A provides an important data point and may constrain models that predict it. The flux upper limits set in Section 2.2 correspond to an intrinsic upper limit of $\sim 2.4 \times 10^{47}$ erg s$^{-1}$ for precursor emission on the 0.1 s timescale, $\sim 7.0 \times 10^{46}$ erg s$^{-1}$ for precursor emission on the 1.0 s timescale, and $\sim 2.2 \times 10^{46}$ erg s$^{-1}$ for extended emission on the 10 s timescale.

Magnetospheric interactions in NS binaries have been proposed as a source of nearly isotropic emission preceding the merger (e.g., Hansen & Lyutikov 2001; Metzger & Zivancev 2016). In the context of such models, the non-detection of precursors associated with GW170817 suggests the absence of strong magnetic fields in the last ∼200 s before merger. Hansen & Lyutikov (2001), for instance, predict a luminosity that depends on the magnetic field $B$ as $L \simeq 7.4 \times 10^{45}\ (B/10^{15}\ {\rm G})^2 (a/10^7\ {\rm cm})^{-7}$ erg s$^{-1}$. We can combine this estimate with the least-constraining GBM intrinsic upper limit above and assume a final separation $a = 3 \times 10^6$ cm before disruption. The resulting limit is $B < 8 \times 10^{13}$ G, which is weaker than the magnetic fields of most known magnetars (Olausen & Kaspi 2014). However, the GBM upper limit still lies within the luminosity range of other similar models (Metzger & Zivancev 2016).

Resonant shattering of the NS crust has also been proposed as a source of emission prior to merger, with a maximum time delay of tens of seconds and nearly isotropic angular distribution (Tsang et al. 2012). The luminosity of such precursor emission depends on the crust breaking strain $\epsilon_{\rm b}$ and the emission timescale $\Delta t$ as $L \simeq 7 \times 10^{48} \epsilon_{\rm b}^2/\Delta t$ erg, from which we can derive $\epsilon_{\rm b}^2 \lesssim 10^{-2}(\Delta t/1\ {\rm s})$. Assuming this mechanism took place in GW170817, and taking $\epsilon_{\rm b} = 0.1$ (Horowitz & Kadau 2009), the emission either lasted more than a few seconds or happened below the GBM energy range, i.e., ∼10 keV. There might also be a dependence of the luminosity on the details of the NS EOS, although that is yet to be investigated in detail. Similarly to magnetospheric interaction, however, resonant shattering emission ultimately requires a sufficiently large magnetic field and a simple explanation for the absence of a signal is again the lack of intense magnetic fields prior to merger.

GBM and SPI-ACS observed no temporally extended gamma-ray emission for GRB 170817A. Such emission would be a signature of a long-lived NS remnant powering the SGRB and our flux limits may suggest instead that the remnant is a BH. Metzger et al. (2008) invoke a long-lived millisecond magnetar to explain SGRBs with extended gamma-ray emission (Norris & Bonnell 2006), and millisecond magnetars have also been suggested as possible causes for the plateaus seen in X-ray afterglows of some SGRBs (Rowlinson et al. 2013). The earliest X-ray observation was only performed 50 ks after GRB 170817A (Evans et al. 2017) and hence limits are only set after this time. Future observations may further constrain this scenario, e.g., radio observations on the timescale of a year (Fong et al. 2016).

We encourage the development of quantitative predictions of luminosity as a function of energy, time and physical parameters of the source, as the multiple upcoming joint observations of BNS mergers suggest the possibility of interesting constraints on the pre-merger physics.

## 7. Conclusion

The joint observation of GW170817 and GRB 170817A confirms the association of SGRBs with BNS mergers. With just one joint event, we have set stringent limits on fundamental physics and probed the central engine of SGRBs in ways that have not been possible with EM data alone, demonstrating the importance of multi-messenger astronomy.

The small time offset and independent localizations, though coarse, allowed an unambiguous association of these two events. Because GRB 170817A occurred nearby, an autonomous trigger on-board GBM alerted follow-up observers to the presence of a counterpart to GW170817. At design sensitivity, however, Advanced LIGO and Virgo could in principle detect GW170817 beyond the distance that any active gamma-ray observatory would trigger on a burst like GRB 170817A. Subthreshold searches for SGRBs can extend the gamma-ray horizon and the detection of GRB 170817A provides motivation for further subthreshold search development.

A joint detection at greater distance and for an SGRB with more typical energetics would allow tighter constraints on the temporal offset and the derived inferences. Should NS–BH binaries also be SGRB progenitors, only a joint detection between GW and EM can provide decisive evidence.

In this Letter we propose several explanations for the observed dimness of GRB 170817A. We suggest joint detections should be more common than previously predicted, and future observations of multiple events should enable a study of the populations of mergers and their associated SGRBs, shedding light on the jet geometry and intrinsic brightness distribution. Furthermore, detections with multiple

GW interferometers can provide more stringent constraints on the inclination angles of these systems. The joint detections of SGRBs arising from BNS and NS–BH mergers will constrain the fraction of SGRBs originating from each progenitor class.

The global network of GW detectors and wide-field gamma-ray instruments, such as *Fermi*-GBM and *INTEGRAL*/SPI-ACS, are critical to the future of multi-messenger astronomy in the GW era.

We dedicate this Letter to the memory of Neil Gehrels. His pioneering work in gamma-ray astronomy and his vision for multi-messenger astrophysics were instrumental to our discoveries.

The authors thank the referees for their invaluable comments and feedback, especially in a timely manner. The authors gratefully acknowledge the support of the United States National Science Foundation (NSF) for the construction and operation of the LIGO Laboratory and Advanced LIGO as well as the Science and Technology Facilities Council (STFC) of the United Kingdom, the Max-Planck-Society (MPS), and the State of Niedersachsen/Germany for support of the construction of Advanced LIGO and construction and operation of the GEO600 detector. Additional support for Advanced LIGO was provided by the Australian Research Council. The authors gratefully acknowledge the Italian Istituto Nazionale di Fisica Nucleare (INFN), the French Centre National de la Recherche Scientifique (CNRS) and the Foundation for Fundamental Research on Matter supported by the Netherlands Organisation for Scientific Research, for the construction and operation of the Virgo detector and the creation and support of the EGO consortium. The authors also gratefully acknowledge research support from these agencies as well as by the Council of Scientific and Industrial Research of India, the Department of Science and Technology, India, the Science & Engineering Research Board (SERB), India, the Ministry of Human Resource Development, India, the Spanish Agencia Estatal de Investigación, the Vicepresidència i Conselleria d'Innovació Recerca i Turisme and the Conselleria d'Educació i Universitat del Govern de les Illes Balears, the Conselleria d'Educació Investigació Cultura i Esport de la Generalitat Valenciana, the National Science Centre of Poland, the Swiss National Science Foundation (SNSF), the Russian Foundation for Basic Research, the Russian Science Foundation, the European Commission, the European Regional Development Funds (ERDF), the Royal Society, the Scottish Funding Council, the Scottish Universities Physics Alliance, the Hungarian Scientific Research Fund (OTKA), the Lyon Institute of Origins (LIO), the National Research, Development and Innovation Office Hungary (NKFI), the National Research Foundation of Korea, Industry Canada and the Province of Ontario through the Ministry of Economic Development and Innovation, the Natural Science and Engineering Research Council Canada, the Canadian Institute for Advanced Research, the Brazilian Ministry of Science, Technology, Innovations, and Communications, the International Center for Theoretical Physics South American Institute for Fundamental Research (ICTP-SAIFR), the Research Grants Council of Hong Kong, the National Natural Science Foundation of China (NSFC), the Leverhulme Trust, the Research Corporation, the Ministry of Science and Technology (*MOST*), Taiwan and the Kavli Foundation. The authors gratefully acknowledge the support of the NSF, STFC, MPS, INFN, CNRS and the State of Niedersachsen/Germany for provision of computational resources.

The USRA co-authors gratefully acknowledge NASA funding through contract NNM13AA43C. The UAH co-authors gratefully acknowledge NASA funding from co-operative agreement NNM11AA01A. E.B. and T.D.C. are supported by an appointment to the NASA Postdoctoral Program at the Goddard Space Flight Center, administered by Universities Space Research Association under contract with NASA. D.K., C.A.W.H., C.M.H., and T.L. gratefully acknowledge NASA funding through the *Fermi* GBM project. Support for the German contribution to GBM was provided by the Bundesministerium für Bildung und Forschung (BMBF) via the Deutsches Zentrum für Luft und Raumfahrt (DLR) under contract number 50 QV 0301. A.v.K. was supported by the Bundesministeriums fr Wirtschaft und Technologie (BMWi) through DLR grant 50 OG 1101. N.C. and J.B. acknowledge support from NSF under grant PHY-1505373. S.M.B. acknowledges support from Science Foundation Ireland under grant 12/IP/1288.

This work is based on observations with *INTEGRAL*, an ESA project with instruments and science data center funded by ESA member states (especially the PI countries: Denmark, France, Germany, Italy, Switzerland, Spain), and with the participation of Russia and the USA. The *INTEGRAL* SPI project has been completed under the responsibility and leadership of CNES. The SPI-ACS detector system has been provided by MPE Garching/Germany. The SPI team is grateful to ASI, CEA, CNES, DLR, ESA, INTA, NASA and OSTC for their support. The Italian *INTEGRAL* team acknowledges the support of ASI/INAF agreement n. 2013-025-R.1. R.D. and A.v.K. acknowledge the German *INTEGRAL* support through DLR grant 50 OG 1101. A.L. and R.S. acknowledge the support from the Russian Science Foundation (grant 14-22-00271). A.D. is funded by Spanish MINECO/FEDER grant ESP2015-65712-C5-1-R. Some of the results in this paper have been derived using the HEALPix (Gorski et al. 2005) package. We are grateful VirtualData from LABEX P2IO for enabling access to the StratusLab academic cloud. We acknowledge the continuous support by the *INTEGRAL* Users Group and the exceptionally efficient support by the teams at ESAC and ESOC for the scheduling of the targeted follow-up observations.

## Appendix A
## Full Derivation of Photospheric Radius

In order to explain the observed soft BB component, we present a model that simultaneously yields predictions on (1) the time difference between the GW emission and the beginning of the gamma-ray radiation, (2) the estimated temperature of the BB component ($k_{\rm B}T_{\rm BB}^{\rm obs} = (10.3 \pm 1.5)$ keV) and (3) its average isotropic luminosity, $L_{\rm iso,BB} = E_{\rm iso,BB}/\Delta t_{\rm BB} = 1.1 \times 10^{46}$ erg s$^{-1}$, where we have used the source interval width of the soft tail spectral fit of $\Delta t_{\rm BB} = 1.15$ s.

Following MR00, our model depends upon three main parameters to provide predictions of the aforementioned three observable quantities. First, the radius from which the fireball is initiated, which is assumed to be $R_0 = 6GM_{\rm BH}/c^2$ (the innermost stable circular orbit of a Schwarzschild BH with a mass equal to $M_{\rm BH}$). In our case $R_0 = 2.5 \times 10^6$ cm, and $M_{\rm BH} = 2.8\,M_\odot$. From numerical models of GRB jets produced in BNS merger remnants (e.g., Aloy et al. 2005), the value of $R_0$ can be associated to the stagnation point of a relativistic outflow, and it is fairly well constrained to be a few gravitational radii of the BH. Second, the initial luminosity of the fireball, $L_0$. This is a free parameter of the model and we note it can be much larger than $L_{\rm iso,BB}$, since the fireball must expand from its initial volume ($\sim R_0^3$) to the size where the photosphere appears (see below). This is also needed, since the observed gamma-ray luminosity, will be a fraction $0 < \epsilon_r \leqslant 1$ of the total (kinetic) luminosity. Finally, the third parameter is the dimensionless entropy of the fireball, $\eta$. For a fireball baryon load $\dot{M}$ and a luminosity $L_0$, $\eta = L/(\dot{M}c^2)$. Typical values of $\eta$ are larger than 100 to prevent the compactness problem (Goodman 1986), which is not an issue for GRB 170817A owing to the lack of emission detected above 511 keV. Therefore, our model may allow for values of $\eta$ substantially smaller than $\sim$100. As we shall see, a combination of $L_0 \simeq 10^{50}$ erg and $\eta \simeq 18$, results in a viable model to account for the delay of GRB with respect to the GW signal and the average luminosity of the soft BB component.

The initial BB temperature in units of the electron rest mass is given by (MR00, Equation (5))

$$\Theta_0 = \left(\frac{k_{\rm B}}{m_e c^2}\right)\left(\frac{L_0}{4\pi R_0^2 c a_r}\right)^{1/4} \simeq 1.5\left(\frac{L_0}{10^{50}\ {\rm erg\ s^{-1}}}\right)^{1/4}\left(\frac{R_0}{2.5 \times 10^6\ {\rm cm}}\right)^{-1/2}, \quad (22)$$

where $a_r = 7.57 \times 10^{-18}$ kg cm$^{-1}$ s$^{-2}$ K$^{-4}$ is the radiation constant. The value of $\Theta_0$ corresponds to a comoving temperature $k_{\rm B}T_0 \simeq 750$ keV. The radius at which the internal energy of the fireball is converted into kinetic energy, i.e., the saturation radius is

$$R_{\rm s} = \eta R_0 \simeq 4.5 \times 10^7\ {\rm cm}\left(\frac{\eta}{18}\right)\left(\frac{R_0}{2.5 \times 10^6\ {\rm cm}}\right). \quad (23)$$

The critical baryon load $\eta_*$ below which the photosphere of an expanding fireball happens after the fireball coasts at constant Lorentz factor $\gamma \simeq \eta$, i.e., at radii larger than $R_{\rm s}$, is given by (MR00)

$$\eta_* = \left(\frac{L_0 \sigma_T Y}{4\pi m_{\rm p} c^3 R_0}\right)^{1/4}, \quad (24)$$

where $Y$ is the number of electrons per baryon, $\sigma_T = 6.65 \times 10^{-25}$ cm$^2$ is the Thomson cross-section. Using fiducial values, we obtain

$$\eta_* \simeq 470\left(\frac{L_0}{10^{50}\ {\rm erg\ s^{-1}}}\right)^{1/4}\left(\frac{R_0}{2.5 \times 10^6\ {\rm cm}}\right)^{-1/4} Y^{1/4}. \quad (25)$$

In the previous expression, we have taken $Y \simeq 1$, which is appropriate once pairs are not present in the system. This is the case for radii larger than $R_{\rm p}$ (MR00)

$$R_{\rm p} = R_0 \frac{\Theta_0}{\Theta_p} \simeq 1.1 \times 10^8\ {\rm cm}\left(\frac{L_0}{10^{50}\ {\rm erg\ s^{-1}}}\right)^{1/4} \times \left(\frac{R_0}{2.5 \times 10^6\ {\rm cm}}\right)^{1/2}\left(\frac{\Theta_p}{0.03}\right)^{-1}, \quad (26)$$

where the comoving dimensionless temperature below which $e^\pm$ pairs drop out of equilibrium is $\Theta_p \simeq 0.03 \simeq 17$ keV. Since we have set $\eta < \eta_*$, the photosphere will happen at a radius (MR00)

$$R_{\rm ph} = \frac{L \sigma_T Y}{4\pi m_{\rm p} c^3 \eta^3} \simeq 2 \times 10^{13}\ {\rm cm}\left(\frac{L_0}{10^{50}\ {\rm erg\ s^{-1}}}\right)\left(\frac{\eta}{18}\right)^{-3} Y. \quad (27)$$

Note that $R_{\rm p} \ll R_{\rm ph}$ for our choice of tunable parameters.

The laboratory frame time needed for the fireball expanding at roughly the speed of light to reach the transparency radius is $t_{\rm ph} \simeq R_{\rm ph}/c$, thus,

$$t_{\rm ph} \simeq 672\ {\rm s}\left(\frac{L_0}{10^{50}\ {\rm erg\ s^{-1}}}\right)\left(\frac{\eta}{18}\right)^{-3} Y. \quad (28)$$

To compute the time delay between a photon emitted at $R_0$ at $t = 0$ (namely, signaling the GW detection) and another one at $R_{\rm ph}$, we must consider that the fireball begins its expansion from rest, in which case we shall apply the following relation between the arrival time and the time at which the photosphere appears

$$t_a \simeq t_{\rm ph}\left(\frac{1}{2\eta^2} + 1 - \cos\vartheta\right), \quad (29)$$

where $\vartheta$ is the angle between the radial direction and the line of sight and $\cos\vartheta$ takes values only in the interval $[v/c, 1]$ (Bianco et al. 2001). Using fiducial values for $\eta$, we obtain

$$t_a^{(1)} \simeq \frac{t_{\rm ph}}{2\gamma^2} \simeq 1.0\ {\rm s}\left(\frac{L_0}{10^{50}\ {\rm erg\ s^{-1}}}\right)\left(\frac{\eta}{18}\right)^{-5} Y, \quad \text{for } \cos\vartheta = 1 \quad (30)$$

$$t_a^{(2)} \simeq \frac{t_{\rm ph}}{\gamma^2} \simeq 2.1\ {\rm s}\left(\frac{L_0}{10^{50}\ {\rm erg\ s^{-1}}}\right)\left(\frac{\eta}{18}\right)^{-5} Y, \quad \text{for } \cos\vartheta = v/c. \quad (31)$$

These values account for a significant fraction of the time delay between the GW and the prompt radiation in the soft thermal peak.

We can now estimate the comoving temperature of the expanding fireball at the photospheric radius (MR00)

$$k_B T_{ph} = k_B T_0 \left(\frac{R_{ph}}{R_s}\right)^{-2/3} \quad (32)$$

$$\simeq 0.13\ \mathrm{keV} \left(\frac{L_0}{10^{50}\ \mathrm{erg\ s^{-1}}}\right)^{-5/12} \left(\frac{R_0}{2.5 \times 10^6\ \mathrm{cm}}\right)^{-5/6} \times \left(\frac{\eta}{18}\right)^{8/3} Y^{-2/3}. \quad (33)$$

This corresponds to an observed temperature, $T_{ph}^{obs} = \eta T_{ph}$

$$k_B T_{ph}^{obs} = \eta k_B T_{ph} \simeq 2.3\ \mathrm{keV} \left(\frac{L_0}{10^{50}\ \mathrm{erg\ s^{-1}}}\right)^{-5/12} \times \left(\frac{R_0}{2.5 \times 10^6\ \mathrm{cm}}\right)^{-5/6} \left(\frac{\eta}{18}\right)^{11/3} Y^{-2/3}. \quad (34)$$

The corresponding BB luminosity is (MR00),

$$L_{ph} = L_0 \left(\frac{R_{ph}}{R_s}\right)^{-2/3} \quad (35)$$

$$\simeq 1.7 \times 10^{46}\ \mathrm{erg\ s^{-1}} \left(\frac{L_0}{10^{50}\ \mathrm{erg\ s^{-1}}}\right)^{7/12} \times \left(\frac{R_0}{2.5 \times 10^6\ \mathrm{cm}}\right)^{-5/6} \left(\frac{\eta}{18}\right)^{8/3} Y^{-2/3}. \quad (36)$$

Thus, with the reference values of our model for $L_0$ and $\eta$, we obtain $L_{ph} \simeq L_{iso,BB}$. On the other hand, $T_{ph}^{obs}$ underpredicts $T_{BB}^{obs}$ by a factor $\sim$4. Nonetheless, we are neglecting Comptonization effects, which may slightly raise the estimated temperature.

## Appendix B
## List of SGRBs with Associated Redshift

Table 2 is a list of possible redshifts for GRBs that have been argued to belong to the short class. The asterisks show the "gold sample" selection with standard cuts on duration ($T_{90} < 2$ s) and localization uncertainty ($\sim$arcsecond or better). The others include SGRBs with extended gamma-ray emission, those slightly longer than 2 s that are spectrally hard or show negligible spectral lag, and bursts that are best localized by *Swift* BAT (so a chance association is possible). Most of these redshifts come from Lien et al. (2016); Siellez et al. (2016), and Fong et al. (2015); for the original citations see references therein. They also include bursts from Tunnicliffe et al. (2014). For these bursts, and those best localized by *Swift* BAT, an individual nearby galaxy may be a chance alignment, but it is statistically unlikely that most of them are false associations.

**Table 2**
Probable SGRBs with Measured Redshifts

| GRB | Any Claimed Redshift | Comment |
|---|---|---|
| 161104A* | 0.788 | |
| 160821B* | 0.16 | |
| 160624A* | 0.483 | |
| 150423A* | 1.394 | |
| 150120A* | 0.46 | |
| 150101B* | 0.134 | |
| 141212A* | 0.596 | |
| 140903A* | 0.351 | |
| 140622A* | 0.959 | |
| 131004A* | 0.717 | |
| 130603B* | 0.356 | |
| 120804A* | 1.3 | |
| 111117A* | 2.211 | Updated value from Selsing et al. (2017) |
| 111020A | 0.019 | Tunnicliffe et al. (2014) |
| 101219A* | 0.718 | |
| 100724A* | 1.288 | |
| 100628A* | 0.102 | |
| 100625A* | 0.452 | |
| 100206A* | 0.407 | |
| 100117A* | 0.915 | |
| 090927 | 1.37 | Arcminute Localization |
| 090515* | 0.403 | |
| 090510 | 0.903 | *Swift* BAT $T_{90} > 2$ s |
| 090426* | 2.609 | |
| 090417A | 0.088 | Arcminute Localization |
| 080905A* | 0.122 | |
| 071227 | 0.381 | SGRB w/Extended Emission |
| 070923 | 0.076 | Arcminute Localization |
| 070809* | 0.473 | Tunnicliffe et al. (2014) mention of possible host at $z = 0.047$ |
| 070729* | 0.8 | |
| 070724A* | 0.457 | |
| 070714B | 0.923 | SGRB w/Extended Emission |
| 070429B* | 0.902 | |
| 061217* | 0.827 | |
| 061210 | 0.41 | SGRB w/Extended Emission |
| 061201* | 0.111 | |
| 061006 | 0.438 | SGRB w/Extended Emission |
| 060801* | 1.13 | |
| 060502B* | 0.287 | |
| 051221A* | 0.546 | |
| 050906 | 0.032 | Tunnicliffe et al. (2014) |
| 050813* | 0.72/1.8 | It is unclear which redshift is correct |
| 050724A | 0.257 | SGRB w/ Extended Emission |
| 050709 | 0.161 | Not detected by *Swift* BAT |
| 050509B* | 0.225 | |

B. P. Abbott[1], R. Abbott[1], T. D. Abbott[2], F. Acernese[3,4], K. Ackley[5,6], C. Adams[7], T. Adams[8], P. Addesso[9], R. X. Adhikari[1], V. B. Adya[10], C. Affeldt[10], M. Afrough[11], B. Agarwal[12], M. Agathos[13], K. Agatsuma[14], N. Aggarwal[15], O. D. Aguiar[16], L. Aiello[17,18], A. Ain[19], P. Ajith[20], B. Allen[10,21,22], G. Allen[12], A. Allocca[23,24], M. A. Aloy[25], P. A. Altin[26], A. Amato[27], A. Ananyeva[1], S. B. Anderson[1], W. G. Anderson[21], S. V. Angelova[28], S. Antier[29], S. Appert[1], K. Arai[1], M. C. Araya[1], J. S. Areeda[30], N. Arnaud[29,31], K. G. Arun[32], S. Ascenzi[33,34], G. Ashton[10], M. Ast[35], S. M. Aston[7], P. Astone[36], D. V. Atallah[37], P. Aufmuth[22], C. Aulbert[10], K. AultONeal[38], C. Austin[2], A. Avila-Alvarez[30], S. Babak[39], P. Bacon[40], M. K. M. Bader[14], S. Bae[41], P. T. Baker[42], F. Baldaccini[43,44], G. Ballardin[31], S. W. Ballmer[45], S. Banagiri[46], J. C. Barayoga[1], S. E. Barclay[47], B. C. Barish[1], D. Barker[48], K. Barkett[49], F. Barone[3,4], B. Barr[47], L. Barsotti[15], M. Barsuglia[40], D. Barta[50], J. Bartlett[48], I. Bartos[5,51], R. Bassiri[52], A. Basti[23,24], J. C. Batch[48], M. Bawaj[44,53], J. C. Bayley[47], M. Bazzan[54,55], B. Bécsy[56], C. Beer[10], M. Bejger[57], I. Belahcene[29], A. S. Bell[47], B. K. Berger[1], G. Bergmann[10], J. J. Bero[58], C. P. L. Berry[59], D. Bersanetti[60], A. Bertolini[14], J. Betzwieser[7], S. Bhagwat[45], R. Bhandare[61], I. A. Bilenko[62], G. Billingsley[1], C. R. Billman[5], J. Birch[7], R. Birney[63], O. Birnholtz[10], S. Biscans[1,15], S. Biscoveanu[6,64], A. Bisht[22], M. Bitossi[24,31], C. Biwer[45], M. A. Bizouard[29], J. K. Blackburn[1], J. Blackman[49], C. D. Blair[1,65], D. G. Blair[65], R. M. Blair[48], S. Bloemen[66], O. Bock[10], N. Bode[10], M. Boer[67], G. Bogaert[67], A. Bohe[39], F. Bondu[68], E. Bonilla[52], R. Bonnand[8], B. A. Boom[14], R. Bork[1], V. Boschi[24,31], S. Bose[19,69], K. Bossie[7], Y. Bouffanais[40], A. Bozzi[31], C. Bradaschia[24], P. R. Brady[21], M. Branchesi[17,18], J. E. Brau[70], T. Briant[71], A. Brillet[67], M. Brinkmann[10], V. Brisson[29], P. Brockill[21], J. E. Broida[72], A. F. Brooks[1], D. A. Brown[45], D. D. Brown[73], S. Brunett[1], C. C. Buchanan[2], A. Buikema[15], T. Bulik[74], H. J. Bulten[14,75], A. Buonanno[39,76], D. Buskulic[8], C. Buy[40], R. L. Byer[52], M. Cabero[10], L. Cadonati[77], G. Cagnoli[27,78], C. Cahillane[1], J. Calderón Bustillo[77], T. A. Callister[1], E. Calloni[4,79], J. B. Camp[80], M. Canepa[60,81], P. Canizares[66], K. C. Cannon[82], H. Cao[73], J. Cao[83], C. D. Capano[10], E. Capocasa[40], F. Carbognani[31], S. Caride[84], M. F. Carney[85], J. Casanueva Diaz[29], C. Casentini[33,34], S. Caudill[14,21], M. Cavaglià[11], F. Cavalier[29], R. Cavalieri[31], G. Cella[24], C. B. Cepeda[1], P. Cerdá-Durán[25], G. Cerretani[23,24], E. Cesarini[34,86], S. J. Chamberlin[64], M. Chan[47], S. Chao[87], P. Charlton[88], E. Chase[89], E. Chassande-Mottin[40], D. Chatterjee[21], K. Chatziioannou[90], B. D. Cheeseboro[42], H. Y. Chen[91], X. Chen[65], Y. Chen[49], H.-P. Cheng[5], H. Chia[5], A. Chincarini[60],

A. Chiummo[31], T. Chmiel[85], H. S. Cho[92], M. Cho[76], J. H. Chow[26], N. Christensen[67,72], Q. Chu[65], A. J. K. Chua[13], S. Chua[71], A. K. W. Chung[93], S. Chung[65], G. Ciani[5,54,55], R. Ciolfi[94,95], C. E. Cirelli[52], A. Cirone[60,81], F. Clara[48], J. A. Clark[77], P. Clearwater[96], F. Cleva[67], C. Cocchieri[11], E. Coccia[17,18], P.-F. Cohadon[71], D. Cohen[29], A. Colla[36,97], C. G. Collette[98], L. R. Cominsky[99], M. Constancio Jr.[16], L. Conti[55], S. J. Cooper[59], P. Corban[7], T. R. Corbitt[2], I. Cordero-Carrión[100], K. R. Corley[51], N. Cornish[101], A. Corsi[84], S. Cortese[31], C. A. Costa[16], M. W. Coughlin[1,72], S. B. Coughlin[89], J.-P. Coulon[67], S. T. Countryman[51], P. Couvares[1], P. B. Covas[102], E. E. Cowan[77], D. M. Coward[65], M. J. Cowart[7], D. C. Coyne[1], R. Coyne[84], J. D. E. Creighton[21], T. D. Creighton[103], J. Cripe[2], S. G. Crowder[104], T. J. Cullen[2,30], A. Cumming[47], L. Cunningham[47], E. Cuoco[31], T. Dal Canton[80], G. Dálya[56], S. L. Danilishin[10,22], S. D'Antonio[34], K. Danzmann[10,22], A. Dasgupta[105], C. F. Da Silva Costa[5], V. Dattilo[31], I. Dave[61], M. Davier[29], D. Davis[45], E. J. Daw[106], B. Day[77], S. De[45], D. DeBra[52], J. Degallaix[27], M. De Laurentis[4,17], S. Deléglise[71], W. Del Pozzo[23,24,59], N. Demos[15], T. Denker[10], T. Dent[10], R. De Pietri[107,108], V. Dergachev[39], R. De Rosa[4,79], R. T. DeRosa[7], C. De Rossi[27,31], R. DeSalvo[109], O. de Varona[10], J. Devenson[28], S. Dhurandhar[19], M. C. Díaz[103], L. Di Fiore[4], M. Di Giovanni[95,110], T. Di Girolamo[4,51,79], A. Di Lieto[23,24], S. Di Pace[36,97], I. Di Palma[36,97], F. Di Renzo[23,24], Z. Doctor[91], V. Dolique[27], F. Donovan[15], K. L. Dooley[11], S. Doravari[10], I. Dorrington[37], R. Douglas[47], M. Dovale Álvarez[59], T. P. Downes[21], M. Drago[10], C. Dreissigacker[10], J. C. Driggers[48], Z. Du[83], M. Ducrot[8], P. Dupej[47], S. E. Dwyer[48], T. B. Edo[106], M. C. Edwards[72], A. Effler[7], H.-B. Eggenstein[10,39], P. Ehrens[1], J. Eichholz[1], S. S. Eikenberry[5], R. A. Eisenstein[15], R. C. Essick[15], D. Estevez[8], Z. B. Etienne[42], T. Etzel[1], M. Evans[15], T. M. Evans[7], M. Factourovich[51], V. Fafone[17,33,34], H. Fair[45], S. Fairhurst[37], X. Fan[83], S. Farinon[60], B. Farr[91], W. M. Farr[59], E. J. Fauchon-Jones[37], M. Favata[111], M. Fays[37], C. Fee[85], H. Fehrmann[10], J. Feicht[1], M. M. Fejer[52], A. Fernandez-Galiana[15], I. Ferrante[23,24], E. C. Ferreira[16], F. Ferrini[31], F. Fidecaro[23,24], D. Finstad[45], I. Fiori[31], D. Fiorucci[40], M. Fishbach[91], R. P. Fisher[45], M. Fitz-Axen[46], R. Flaminio[27,112], M. Fletcher[47], H. Fong[90], J. A. Font[25,113], P. W. F. Forsyth[26], S. S. Forsyth[77], J.-D. Fournier[67], S. Frasca[36,97], F. Frasconi[24], Z. Frei[56], A. Freise[59], R. Frey[70], V. Frey[29], E. M. Fries[1], P. Fritschel[15], V. V. Frolov[7], P. Fulda[5], M. Fyffe[7], H. Gabbard[47], B. U. Gadre[19], S. M. Gaebel[59], J. R. Gair[114], L. Gammaitoni[43], M. R. Ganija[73], S. G. Gaonkar[19], C. Garcia-Quiros[102], F. Garufi[4,79], B. Gateley[48], S. Gaudio[38], G. Gaur[115], V. Gayathri[116], N. Gehrels[80,†], G. Gemme[60], E. Genin[31], A. Gennai[24], D. George[12], J. George[61], L. Gergely[117], V. Germain[8], S. Ghonge[77], Abhirup Ghosh[20], Archisman Ghosh[14,20], S. Ghosh[14,21,66], J. A. Giaime[2,7], K. D. Giardina[7], A. Giazotto[24], K. Gill[38], L. Glover[109], E. Goetz[118], R. Goetz[5], S. Gomes[37], B. Goncharov[6], G. González[2], J. M. Gonzalez Castro[23,24], A. Gopakumar[119], M. L. Gorodetsky[62], S. E. Gossan[1], M. Gosselin[31], R. Gouaty[8], A. Grado[4,120], C. Graef[47], M. Granata[27], A. Grant[47], S. Gras[15], C. Gray[48], G. Greco[121,122], A. C. Green[59], E. M. Gretarsson[38], P. Groot[66], H. Grote[10], S. Grunewald[39], P. Gruning[29], G. M. Guidi[121,122], X. Guo[83], A. Gupta[64], M. K. Gupta[105], K. E. Gushwa[1], E. K. Gustafson[1], R. Gustafson[118], O. Halim[17,18], B. R. Hall[69], E. D. Hall[15], E. Z. Hamilton[37], G. Hammond[47], M. Haney[123], M. M. Hanke[10], J. Hanks[48], C. Hanna[64], M. D. Hannam[37], O. A. Hannuksela[93], J. Hanson[7], T. Hardwick[2], J. Harms[17,18], G. M. Harry[124], I. W. Harry[39], M. J. Hart[47], C.-J. Haster[90], K. Haughian[47], J. Healy[58], A. Heidmann[71], M. C. Heintze[7], H. Heitmann[67], P. Hello[29], G. Hemming[31], M. Hendry[47], I. S. Heng[47], J. Hennig[47], A. W. Heptonstall[1], M. Heurs[10,22], S. Hild[47], T. Hinderer[66], D. Hoak[31], D. Hofman[27], K. Holt[7], D. E. Holz[91], P. Hopkins[37], C. Horst[21], J. Hough[47], E. A. Houston[47], E. J. Howell[65], A. Hreibi[67], Y. M. Hu[10], E. A. Huerta[12], D. Huet[29], B. Hughey[38], S. Husa[102], S. H. Huttner[47], T. Huynh-Dinh[7], N. Indik[10], R. Inta[84], G. Intini[36,97], H. N. Isa[47], J.-M. Isac[71], M. Isi[1], B. R. Iyer[20], K. Izumi[48], T. Jacqmin[71], K. Jani[77], P. Jaranowski[125], S. Jawahar[63], F. Jiménez-Forteza[102], W. W. Johnson[2], N. K. Johnson-McDaniel[13], D. I. Jones[126], R. Jones[47], R. J. G. Jonker[14], L. Ju[65], J. Junker[10], C. V. Kalaghatgi[37], V. Kalogera[89], B. Kamai[1], S. Kandhasamy[7], G. Kang[41], J. B. Kanner[1], S. J. Kapadia[21], S. Karki[70], K. S. Karvinen[10], M. Kasprzack[2], W. Kastaun[10], M. Katolik[12], E. Katsavounidis[15], W. Katzman[7], S. Kaufer[22], K. Kawabe[48], F. Kéfélian[67], D. Keitel[47], A. J. Kemball[12], R. Kennedy[106], C. Kent[37], J. S. Key[127], F. Y. Khalili[62], I. Khan[17,34], S. Khan[10], Z. Khan[105], E. A. Khazanov[128], N. Kijbunchoo[26], Chunglee Kim[129], J. C. Kim[130], K. Kim[93], W. Kim[73], W. S. Kim[131], Y.-M. Kim[92], S. J. Kimbrell[77], E. J. King[73], P. J. King[48], M. Kinley-Hanlon[124], R. Kirchhoff[10], J. S. Kissel[48], L. Kleybolte[35], S. Klimenko[5], T. D. Knowles[42], P. Koch[10], S. M. Koehlenbeck[10], S. Koley[14], V. Kondrashov[1], A. Kontos[15], M. Korobko[35], W. Z. Korth[1], I. Kowalska[74], D. B. Kozak[1], C. Krämer[10], V. Kringel[10], B. Krishnan[10], A. Królak[132,133], G. Kuehn[10], P. Kumar[90], R. Kumar[105], S. Kumar[20], L. Kuo[87], A. Kutynia[132], S. Kwang[21], B. D. Lackey[39], K. H. Lai[93], M. Landry[48], R. N. Lang[134], J. Lange[58], B. Lantz[52], R. K. Lanza[15], A. Lartaux-Vollard[29], P. D. Lasky[6], M. Laxen[7], A. Lazzarini[1], C. Lazzaro[55], P. Leaci[36,97], S. Leavey[47], C. H. Lee[92], H. K. Lee[135], H. M. Lee[136], H. W. Lee[130], K. Lee[47], J. Lehmann[10], A. Lenon[42], M. Leonardi[95,110], N. Leroy[29], N. Letendre[8], Y. Levin[6], T. G. F. Li[93], S. D. Linker[109], T. B. Littenberg[137], J. Liu[65], R. K. L. Lo[93], N. A. Lockerbie[63], L. T. London[37], J. E. Lord[45], M. Lorenzini[17,18], V. Loriette[138], M. Lormand[7], G. Losurdo[24], J. D. Lough[10], C. O. Lousto[58], G. Lovelace[30], H. Lück[10,22], D. Lumaca[33,34], A. P. Lundgren[10], R. Lynch[15], Y. Ma[49], R. Macas[37], S. Macfoy[28], B. Machenschalk[10], M. MacInnis[15], D. M. Macleod[37], I. Magaña Hernandez[21], F. Magaña-Sandoval[45], L. Magaña Zertuche[45], R. M. Magee[64], E. Majorana[36], I. Maksimovic[138], N. Man[67], V. Mandic[46], V. Mangano[47], G. L. Mansell[26], M. Manske[21,26], M. Mantovani[31], F. Marchesoni[44,53], F. Marion[8], S. Márka[51], Z. Márka[51], C. Markakis[12], A. S. Markosyan[52], A. Markowitz[1], E. Maros[1], A. Marquina[100], F. Martelli[121,122], L. Martellini[67], I. W. Martin[47], R. M. Martin[111], D. V. Martynov[15], K. Mason[15], E. Massera[106], A. Masserot[8], T. J. Massinger[1],

M. Masso-Reid[47], S. Mastrogiovanni[36,97], A. Matas[46], F. Matichard[1,15], L. Matone[51], N. Mavalvala[15], N. Mazumder[69], R. McCarthy[48], D. E. McClelland[26], S. McCormick[7], L. McCuller[15], S. C. McGuire[139], G. McIntyre[1], J. McIver[1], D. J. McManus[26], L. McNeill[6], T. McRae[26], S. T. McWilliams[42], D. Meacher[64], G. D. Meadors[10,39], M. Mehmet[10], J. Meidam[14], E. Mejuto-Villa[9], A. Melatos[96], G. Mendell[48], R. A. Mercer[21], E. L. Merilh[48], M. Merzougui[67], S. Meshkov[1], C. Messenger[47], C. Messick[64], R. Metzdorff[71], P. M. Meyers[46], H. Miao[59], C. Michel[27], H. Middleton[59], E. E. Mikhailov[140], L. Milano[4,79], A. L. Miller[5,36,97], B. B. Miller[89], J. Miller[15], M. Millhouse[101], M. C. Milovich-Goff[109], O. Minazzoli[67,141], Y. Minenkov[34], J. Ming[39], C. Mishra[142], S. Mitra[19], V. P. Mitrofanov[62], G. Mitselmakher[5], R. Mittleman[15], D. Moffa[85], A. Moggi[24], K. Mogushi[11], M. Mohan[31], S. R. P. Mohapatra[15], M. Montani[121,122], C. J. Moore[13], D. Moraru[48], G. Moreno[48], S. R. Morriss[103], B. Mours[8], C. M. Mow-Lowry[59], G. Mueller[5], A. W. Muir[37], Arunava Mukherjee[10], D. Mukherjee[21], S. Mukherjee[103], N. Mukund[19], A. Mullavey[7], J. Munch[73], E. A. Muñiz[45], M. Muratore[38], P. G. Murray[47], K. Napier[77], I. Nardecchia[33,34], L. Naticchioni[36,97], R. K. Nayak[143], J. Neilson[109], G. Nelemans[14,66], T. J. N. Nelson[7], M. Nery[10], A. Neunzert[118], L. Nevin[1], J. M. Newport[124], G. Newton[47,‡], K. K. Y. Ng[93], T. T. Nguyen[26], D. Nichols[66], A. B. Nielsen[10], S. Nissanke[14,66], A. Nitz[10], A. Noack[10], F. Nocera[31], D. Nolting[7], C. North[37], L. K. Nuttall[37], J. Oberling[48], G. D. O'Dea[109], G. H. Ogin[144], J. J. Oh[131], S. H. Oh[131], F. Ohme[10], M. A. Okada[16], M. Oliver[102], P. Oppermann[10], Richard J. Oram[7], B. O'Reilly[7], R. Ormiston[46], L. F. Ortega[5], R. O'Shaughnessy[58], S. Ossokine[39], D. J. Ottaway[73], H. Overmier[7], B. J. Owen[84], A. E. Pace[64], J. Page[137], M. A. Page[65], A. Pai[116,145], S. A. Pai[61], J. R. Palamos[70], O. Palashov[128], C. Palomba[36], A. Pal-Singh[35], Howard Pan[87], Huang-Wei Pan[87], B. Pang[49], P. T. H. Pang[93], C. Pankow[89], F. Pannarale[37], B. C. Pant[61], F. Paoletti[24], A. Paoli[31], M. A. Papa[10,21,39], A. Parida[19], W. Parker[7], D. Pascucci[47], A. Pasqualetti[31], R. Passaquieti[23,24], D. Passuello[24], M. Patil[133], B. Patricelli[24,146], B. L. Pearlstone[47], M. Pedraza[1], R. Pedurand[27,147], L. Pekowsky[45], A. Pele[7], S. Penn[148], C. J. Perez[48], A. Perreca[1,95,110], L. M. Perri[89], H. P. Pfeiffer[39,90], M. Phelps[47], O. J. Piccinni[36,97], M. Pichot[67], F. Piergiovanni[121,122], V. Pierro[9], G. Pillant[31], L. Pinard[27], I. M. Pinto[9], M. Pirello[48], M. Pitkin[47], M. Poe[21], R. Poggiani[23,24], P. Popolizio[31], E. K. Porter[40], A. Post[10], J. Powell[47,149], J. Prasad[19], J. W. W. Pratt[38], G. Pratten[102], V. Predoi[37], T. Prestegard[21], M. Prijatelj[10], M. Principe[9], S. Privitera[39], G. A. Prodi[95,110], L. G. Prokhorov[62], O. Puncken[10], M. Punturo[44], P. Puppo[36], M. Pürrer[39], H. Qi[21], V. Quetschke[103], E. A. Quintero[1], R. Quitzow-James[70], F. J. Raab[48], D. S. Rabeling[26], H. Radkins[48], P. Raffai[56], S. Raja[61], C. Rajan[61], B. Rajbhandari[84], M. Rakhmanov[103], K. E. Ramirez[103], A. Ramos-Buades[102], P. Rapagnani[36,97], V. Raymond[39], M. Razzano[23,24], J. Read[30], T. Regimbau[67], L. Rei[60], S. Reid[63], D. H. Reitze[1,5], W. Ren[12], S. D. Reyes[45], F. Ricci[36,97], P. M. Ricker[12], S. Rieger[10], K. Riles[118], M. Rizzo[58], N. A. Robertson[1,47], R. Robie[47], F. Robinet[29], A. Rocchi[34], L. Rolland[8], J. G. Rollins[1], V. J. Roma[70], R. Romano[3,4], C. L. Romel[48], J. H. Romie[7], D. Rosińska[57,150], M. P. Ross[151], S. Rowan[47], A. Rüdiger[10], P. Ruggi[31], G. Rutins[28], K. Ryan[48], S. Sachdev[1], T. Sadecki[48], L. Sadeghian[21], M. Sakellariadou[152], L. Salconi[31], M. Saleem[116], F. Salemi[10], A. Samajdar[143], L. Sammut[6], L. M. Sampson[89], E. J. Sanchez[1], L. E. Sanchez[1], N. Sanchis-Gual[25], V. Sandberg[48], J. R. Sanders[45], B. Sassolas[27], B. S. Sathyaprakash[37,64], P. R. Saulson[45], O. Sauter[118], R. L. Savage[48], A. Sawadsky[35], P. Schale[70], M. Scheel[49], J. Scheuer[89], J. Schmidt[10], P. Schmidt[1,66], R. Schnabel[35], R. M. S. Schofield[70], A. Schönbeck[35], E. Schreiber[10], D. Schuette[10,22], B. W. Schulte[10], B. F. Schutz[10,37], S. G. Schwalbe[38], J. Scott[47], S. M. Scott[26], E. Seidel[12], D. Sellers[7], A. S. Sengupta[153], D. Sentenac[31], V. Sequino[17,33,34], A. Sergeev[128], D. A. Shaddock[26], T. J. Shaffer[48], A. A. Shah[137], M. S. Shahriar[89], M. B. Shaner[109], L. Shao[39], B. Shapiro[52], P. Shawhan[76], A. Sheperd[21], D. H. Shoemaker[15], D. M. Shoemaker[77], K. Siellez[77], X. Siemens[21], M. Sieniawska[57], D. Sigg[48], A. D. Silva[16], L. P. Singer[80], A. Singh[10,22,39], A. Singhal[17,36], A. M. Sintes[102], B. J. J. Slagmolen[26], B. Smith[7], J. R. Smith[30], R. J. E. Smith[1,6], S. Somala[154], E. J. Son[131], J. A. Sonnenberg[21], B. Sorazu[47], F. Sorrentino[60], T. Souradeep[19], A. P. Spencer[47], A. K. Srivastava[105], K. Staats[38], A. Staley[51], M. Steinke[10], J. Steinlechner[35,47], S. Steinlechner[35], D. Steinmeyer[10], S. P. Stevenson[59,149], R. Stone[103], D. J. Stops[59], K. A. Strain[47], G. Stratta[121,122], S. E. Strigin[62], A. Strunk[48], R. Sturani[155], A. L. Stuver[7], T. Z. Summerscales[156], L. Sun[96], S. Sunil[105], J. Suresh[19], P. J. Sutton[37], B. L. Swinkels[31], M. J. Szczepańczyk[38], M. Tacca[14], S. C. Tait[47], C. Talbot[6], D. Talukder[70], D. B. Tanner[5], M. Tápai[117], A. Taracchini[39], J. D. Tasson[72], J. A. Taylor[137], R. Taylor[1], S. V. Tewari[148], T. Theeg[10], F. Thies[10], E. G. Thomas[59], M. Thomas[7], P. Thomas[48], K. A. Thorne[7], K. S. Thorne[49], E. Thrane[6], S. Tiwari[17,95], V. Tiwari[37], K. V. Tokmakov[63], K. Toland[47], M. Tonelli[23,24], Z. Tornasi[47], A. Torres-Forné[25], C. I. Torrie[1], D. Töyrä[59], F. Travasso[31,44], G. Traylor[7], J. Trinastic[5], M. C. Tringali[95,110], L. Trozzo[24,157], K. W. Tsang[14], M. Tse[15], R. Tso[1], L. Tsukada[82], D. Tsuna[82], D. Tuyenbayev[103], K. Ueno[21], D. Ugolini[158], C. S. Unnikrishnan[119], A. L. Urban[1], S. A. Usman[37], H. Vahlbruch[22], G. Vajente[1], G. Valdes[2], N. van Bakel[14], M. van Beuzekom[14], J. F. J. van den Brand[14,75], C. Van Den Broeck[14,159], D. C. Vander-Hyde[45], L. van der Schaaf[14], J. V. van Heijningen[14], A. A. van Veggel[47], M. Vardaro[54,55], V. Varma[49], S. Vass[1], M. Vasúth[50], A. Vecchio[59], G. Vedovato[55], J. Veitch[47], P. J. Veitch[73], K. Venkateswara[151], G. Venugopalan[1], D. Verkindt[8], F. Vetrano[121,122], A. Viceré[121,122], A. D. Viets[21], S. Vinciguerra[59], D. J. Vine[28], J.-Y. Vinet[67], S. Vitale[15], T. Vo[45], H. Vocca[43,44], C. Vorvick[48], S. P. Vyatchanin[62], A. R. Wade[1], L. E. Wade[85], M. Wade[85], R. Walet[14], M. Walker[30], L. Wallace[1], S. Walsh[10,21,39], G. Wang[17,122], H. Wang[59], J. Z. Wang[64], W. H. Wang[103], Y. F. Wang[93], R. L. Ward[26], J. Warner[48], M. Was[8], J. Watchi[98], B. Weaver[48], L.-W. Wei[10,22], M. Weinert[10], A. J. Weinstein[1], R. Weiss[15], L. Wen[65], E. K. Wessel[12], P. Weßels[10], J. Westerweck[10], T. Westphal[10], K. Wette[26], J. T. Whelan[58], S. E. Whitcomb[1], B. F. Whiting[5],

C. Whittle[6], D. Wilken[10], D. Williams[47], R. D. Williams[1], A. R. Williamson[66], J. L. Willis[1,160], B. Willke[10,22], M. H. Wimmer[10], W. Winkler[10], C. C. Wipf[1], H. Wittel[10,22], G. Woan[47], J. Woehler[10], J. Wofford[58], K. W. K. Wong[93], J. Worden[48], J. L. Wright[47], D. S. Wu[10], D. M. Wysocki[58], S. Xiao[1], H. Yamamoto[1], C. C. Yancey[76], L. Yang[161], M. J. Yap[26], M. Yazback[5], Hang Yu[15], Haocun Yu[15], M. Yvert[8], A. Zadrożny[132], M. Zanolin[38], T. Zelenova[31], J.-P. Zendri[55], M. Zevin[89], L. Zhang[1], M. Zhang[140], T. Zhang[47], Y.-H. Zhang[58], C. Zhao[65], M. Zhou[89], Z. Zhou[89], S. J. Zhu[10,39], X. J. Zhu[6], A. B. Zimmerman[90], M. E. Zucker[1,15], J. Zweizig[1],
(LIGO Scientific Collaboration and Virgo Collaboration)

E. Burns[80], P. Veres[162], D. Kocevski[137], J. Racusin[80], A. Goldstein[163], V. Connaughton[163], M. S. Briggs[162,164], L. Blackburn[15,165], R. Hamburg[162,164], C. M. Hui[137], A. von Kienlin[166], J. McEnery[80], R. D. Preece[162,164], C. A. Wilson-Hodge[137], E. Bissaldi[167,168], W. H. Cleveland[163], M. H. Gibby[169], M. M. Giles[169], R. M. Kippen[170], S. McBreen[171], C. A. Meegan[162], W. S. Paciesas[163], S. Poolakkil[162,164], O. J. Roberts[163], M. Stanbro[164],
(*Fermi* Gamma-ray Burst Monitor)

and
V. Savchenko[172], C. Ferrigno[172], E. Kuulkers[173], A. Bazzano[174], E. Bozzo[172], S. Brandt[175], J. Chenevez[175], T. J.-L. Courvoisier[172], R. Diehl[166], A. Domingo[176], L. Hanlon[177], E. Jourdain[178], P. Laurent[40,179], F. Lebrun[180], A. Lutovinov[181], S. Mereghetti[182], L. Natalucci[174], J. Rodi[174], J.-P. Roques[178], R. Sunyaev[181,183], and P. Ubertini[174]
(INTEGRAL)

† Deceased, 2017 February.
‡ Deceased, 2016 December.

1 LIGO, California Institute of Technology, Pasadena, CA 91125, USA
2 Louisiana State University, Baton Rouge, LA 70803, USA
3 Università di Salerno, Fisciano, I-84084 Salerno, Italy
4 INFN, Sezione di Napoli, Complesso Universitario di Monte S.Angelo, I-80126 Napoli, Italy
5 University of Florida, Gainesville, FL 32611, USA
6 OzGrav, School of Physics & Astronomy, Monash University, Clayton, VIC 3800, Australia
7 LIGO Livingston Observatory, Livingston, LA 70754, USA
8 Laboratoire d'Annecy-le-Vieux de Physique des Particules (LAPP), Université Savoie Mont Blanc, CNRS/IN2P3, F-74941 Annecy, France
9 University of Sannio at Benevento, I-82100 Benevento, Italy and INFN, Sezione di Napoli, I-80100 Napoli, Italy
10 Max Planck Institute for Gravitational Physics (Albert Einstein Institute), D-30167 Hannover, Germany
11 The University of Mississippi, University, MS 38677, USA
12 NCSA, University of Illinois at Urbana-Champaign, Urbana, IL 61801, USA
13 University of Cambridge, Cambridge CB2 1TN, UK
14 Nikhef, Science Park, 1098 XG Amsterdam, The Netherlands
15 LIGO, Massachusetts Institute of Technology, Cambridge, MA 02139, USA
16 Instituto Nacional de Pesquisas Espaciais, 12227-010 São José dos Campos, São Paulo, Brazil
17 Gran Sasso Science Institute (GSSI), I-67100 L'Aquila, Italy
18 INFN, Laboratori Nazionali del Gran Sasso, I-67100 Assergi, Italy
19 Inter-University Centre for Astronomy and Astrophysics, Pune 411007, India
20 International Centre for Theoretical Sciences, Tata Institute of Fundamental Research, Bengaluru 560089, India
21 University of Wisconsin-Milwaukee, Milwaukee, WI 53201, USA
22 Leibniz Universität Hannover, D-30167 Hannover, Germany
23 Università di Pisa, I-56127 Pisa, Italy
24 INFN, Sezione di Pisa, I-56127 Pisa, Italy
25 Departamento de Astronomía y Astrofísica, Universitat de València, E-46100 Burjassot, València, Spain
26 OzGrav, Australian National University, Canberra, ACT 0200, Australia
27 Laboratoire des Matériaux Avancés (LMA), CNRS/IN2P3, F-69622 Villeurbanne, France
28 SUPA, University of the West of Scotland, Paisley PA1 2BE, UK
29 LAL, Univ. Paris-Sud, CNRS/IN2P3, Université Paris-Saclay, F-91898 Orsay, France
30 California State University Fullerton, Fullerton, CA 92831, USA
31 European Gravitational Observatory (EGO), I-56021 Cascina, Pisa, Italy
32 Chennai Mathematical Institute, Chennai 603103, India
33 Università di Roma Tor Vergata, I-00133 Roma, Italy
34 INFN, Sezione di Roma Tor Vergata, I-00133 Roma, Italy
35 Universität Hamburg, D-22761 Hamburg, Germany
36 INFN, Sezione di Roma, I-00185 Roma, Italy
37 Cardiff University, Cardiff CF24 3AA, UK
38 Embry-Riddle Aeronautical University, Prescott, AZ 86301, USA
39 Max Planck Institute for Gravitational Physics (Albert Einstein Institute), D-14476 Potsdam-Golm, Germany
40 APC, AstroParticule et Cosmologie, Université Paris Diderot, CNRS/IN2P3, CEA/Irfu, Observatoire de Paris, Sorbonne Paris Cité, F-75205 Paris Cedex 13, France
41 Korea Institute of Science and Technology Information, Daejeon 34141, Korea
42 West Virginia University, Morgantown, WV 26506, USA
43 Università di Perugia, I-06123 Perugia, Italy

[44] INFN, Sezione di Perugia, I-06123 Perugia, Italy
[45] Syracuse University, Syracuse, NY 13244, USA
[46] University of Minnesota, Minneapolis, MN 55455, USA
[47] SUPA, University of Glasgow, Glasgow G12 8QQ, UK
[48] LIGO Hanford Observatory, Richland, WA 99352, USA
[49] Caltech CaRT, Pasadena, CA 91125, USA
[50] Wigner RCP, RMKI, H-1121 Budapest, Konkoly Thege Miklós út 29-33, Hungary
[51] Columbia University, New York, NY 10027, USA
[52] Stanford University, Stanford, CA 94305, USA
[53] Università di Camerino, Dipartimento di Fisica, I-62032 Camerino, Italy
[54] Università di Padova, Dipartimento di Fisica e Astronomia, I-35131 Padova, Italy
[55] INFN, Sezione di Padova, I-35131 Padova, Italy
[56] Institute of Physics, Eötvös University, Pázmány P. s. 1/A, Budapest 1117, Hungary
[57] Nicolaus Copernicus Astronomical Center, Polish Academy of Sciences, 00-716, Warsaw, Poland
[58] Rochester Institute of Technology, Rochester, NY 14623, USA
[59] University of Birmingham, Birmingham B15 2TT, UK
[60] INFN, Sezione di Genova, I-16146 Genova, Italy
[61] RRCAT, Indore MP 452013, India
[62] Faculty of Physics, Lomonosov Moscow State University, Moscow 119991, Russia
[63] SUPA, University of Strathclyde, Glasgow G1 1XQ, UK
[64] The Pennsylvania State University, University Park, PA 16802, USA
[65] OzGrav, University of Western Australia, Crawley, WA 6009, Australia
[66] Department of Astrophysics/IMAPP, Radboud University Nijmegen, P.O. Box 9010, 6500 GL Nijmegen, The Netherlands
[67] Artemis, Université Côte d'Azur, Observatoire Côte d'Azur, CNRS, CS 34229, F-06304 Nice Cedex 4, France
[68] Institut FOTON, CNRS, Université de Rennes 1, F-35042 Rennes, France
[69] Washington State University, Pullman, WA 99164, USA
[70] University of Oregon, Eugene, OR 97403, USA
[71] Laboratoire Kastler Brossel, UPMC-Sorbonne Universités, CNRS, ENS-PSL Research University, Collège de France, F-75005 Paris, France
[72] Carleton College, Northfield, MN 55057, USA
[73] OzGrav, University of Adelaide, Adelaide, SA 5005, Australia
[74] Astronomical Observatory Warsaw University, 00-478 Warsaw, Poland
[75] VU University Amsterdam, 1081 HV Amsterdam, The Netherlands
[76] University of Maryland, College Park, MD 20742, USA
[77] Center for Relativistic Astrophysics, Georgia Institute of Technology, Atlanta, GA 30332, USA
[78] Université Claude Bernard Lyon 1, F-69622 Villeurbanne, France
[79] Università di Napoli 'Federico II,'Complesso Universitario di Monte S.Angelo, I-80126 Napoli, Italy
[80] NASA Goddard Space Flight Center, Greenbelt, MD 20771, USA
[81] Dipartimento di Fisica, Università degli Studi di Genova, I-16146 Genova, Italy
[82] RESCEU, University of Tokyo, Tokyo, 113-0033, Japan
[83] Tsinghua University, Beijing 100084, China
[84] Texas Tech University, Lubbock, TX 79409, USA
[85] Kenyon College, Gambier, OH 43022, USA
[86] Museo Storico della Fisica e Centro Studi e Ricerche Enrico Fermi, I-00184 Roma, Italy
[87] National Tsing Hua University, Hsinchu City, 30013 Taiwan, Republic of China
[88] Charles Sturt University, Wagga Wagga, NSW 2678, Australia
[89] Center for Interdisciplinary Exploration & Research in Astrophysics (CIERA), Northwestern University, Evanston, IL 60208, USA
[90] Canadian Institute for Theoretical Astrophysics, University of Toronto, Toronto, ON M5S 3H8, Canada
[91] University of Chicago, Chicago, IL 60637, USA
[92] Pusan National University, Busan 46241, Korea
[93] The Chinese University of Hong Kong, Shatin, NT, Hong Kong
[94] INAF, Osservatorio Astronomico di Padova, I-35122 Padova, Italy
[95] INFN, Trento Institute for Fundamental Physics and Applications, I-38123 Povo, Trento, Italy
[96] OzGrav, University of Melbourne, Parkville, VIC 3010, Australia
[97] Università di Roma "La Sapienza," I-00185 Roma, Italy
[98] Université Libre de Bruxelles, Brussels B-1050, Belgium
[99] Sonoma State University, Rohnert Park, CA 94928, USA
[100] Departamento de Matemáticas, Universitat de València, E-46100 Burjassot, València, Spain
[101] Montana State University, Bozeman, MT 59717, USA
[102] Universitat de les Illes Balears, IAC3—IEEC, E-07122 Palma de Mallorca, Spain
[103] The University of Texas Rio Grande Valley, Brownsville, TX 78520, USA
[104] Bellevue College, Bellevue, WA 98007, USA
[105] Institute for Plasma Research, Bhat, Gandhinagar 382428, India
[106] The University of Sheffield, Sheffield S10 2TN, UK
[107] Dipartimento di Scienze Matematiche, Fisiche e Informatiche, Università di Parma, I-43124 Parma, Italy
[108] INFN, Sezione di Milano Bicocca, Gruppo Collegato di Parma, I-43124 Parma, Italy
[109] California State University, Los Angeles, 5151 State University Dr, Los Angeles, CA 90032, USA
[110] Università di Trento, Dipartimento di Fisica, I-38123 Povo, Trento, Italy
[111] Montclair State University, Montclair, NJ 07043, USA
[112] National Astronomical Observatory of Japan, 2-21-1 Osawa, Mitaka, Tokyo 181-8588, Japan
[113] Observatori Astronòmic, Universitat de València, E-46980 Paterna, València, Spain
[114] School of Mathematics, University of Edinburgh, Edinburgh EH9 3FD, UK
[115] University and Institute of Advanced Research, Koba Institutional Area, Gandhinagar Gujarat 382007, India
[116] IISER-TVM, CET Campus, Trivandrum Kerala 695016, India
[117] University of Szeged, Dóm tér 9, Szeged 6720, Hungary
[118] University of Michigan, Ann Arbor, MI 48109, USA

[119] Tata Institute of Fundamental Research, Mumbai 400005, India
[120] INAF, Osservatorio Astronomico di Capodimonte, I-80131 Napoli, Italy
[121] Università degli Studi di Urbino "Carlo Bo," I-61029 Urbino, Italy
[122] INFN, Sezione di Firenze, I-50019 Sesto Fiorentino, Firenze, Italy
[123] Physik-Institut, University of Zurich, Winterthurerstrasse 190, 8057 Zurich, Switzerland
[124] American University, Washington, DC 20016, USA
[125] University of Białystok, 15-424 Białystok, Poland
[126] University of Southampton, Southampton SO17 1BJ, UK
[127] University of Washington Bothell, 18115 Campus Way NE, Bothell, WA 98011, USA
[128] Institute of Applied Physics, Nizhny Novgorod, 603950, Russia
[129] Korea Astronomy and Space Science Institute, Daejeon 34055, Korea
[130] Inje University Gimhae, South Gyeongsang 50834, Korea
[131] National Institute for Mathematical Sciences, Daejeon 34047, Korea
[132] NCBJ, 05-400 Świerk-Otwock, Poland
[133] Institute of Mathematics, Polish Academy of Sciences, 00656 Warsaw, Poland
[134] Hillsdale College, Hillsdale, MI 49242, USA
[135] Hanyang University, Seoul 04763, Korea
[136] Seoul National University, Seoul 08826, Korea
[137] NASA Marshall Space Flight Center, Huntsville, AL 35812, USA
[138] ESPCI, CNRS, F-75005 Paris, France
[139] Southern University and A&M College, Baton Rouge, LA 70813, USA
[140] College of William and Mary, Williamsburg, VA 23187, USA
[141] Centre Scientifique de Monaco, 8 quai Antoine Ier, MC-98000, Monaco
[142] Indian Institute of Technology Madras, Chennai 600036, India
[143] IISER-Kolkata, Mohanpur, West Bengal 741252, India
[144] Whitman College, 345 Boyer Avenue, Walla Walla, WA 99362, USA
[145] Indian Institute of Technology Bombay, Powai, Mumbai, Maharashtra 400076, India
[146] Scuola Normale Superiore, Piazza dei Cavalieri 7, I-56126 Pisa, Italy
[147] Université de Lyon, F-69361 Lyon, France
[148] Hobart and William Smith Colleges, Geneva, NY 14456, USA
[149] OzGrav, Swinburne University of Technology, Hawthorn VIC 3122, Australia
[150] Janusz Gil Institute of Astronomy, University of Zielona Góra, 65-265 Zielona Góra, Poland
[151] University of Washington, Seattle, WA 98195, USA
[152] King's College London, University of London, London WC2R 2LS, UK
[153] Indian Institute of Technology, Gandhinagar Ahmedabad Gujarat 382424, India
[154] Indian Institute of Technology Hyderabad, Sangareddy, Khandi, Telangana 502285, India
[155] International Institute of Physics, Universidade Federal do Rio Grande do Norte, Natal RN 59078-970, Brazil
[156] Andrews University, Berrien Springs, MI 49104, USA
[157] Università di Siena, I-53100 Siena, Italy
[158] Trinity University, San Antonio, TX 78212, USA
[159] Van Swinderen Institute for Particle Physics and Gravity, University of Groningen, Nijenborgh 4, 9747 AG Groningen, The Netherlands
[160] Abilene Christian University, Abilene, TX 79699, USA
[161] Colorado State University, Fort Collins, CO 80523, USA
[162] Center for Space Plasma and Aeronomic Research, University of Alabama in Huntsville, 320 Sparkman Drive, Huntsville, AL 35899, USA
[163] Science and Technology Institute, Universities Space Research Association, Huntsville, AL 35805, USA
[164] Space Science Department, University of Alabama in Huntsville, 320 Sparkman Drive, Huntsville, AL 35899, USA
[165] Harvard-Smithsonian Center for Astrophysics, 60 Garden St, Cambridge, MA 02138, USA
[166] Max-Planck-Institut für extraterrestrische Physik, Giessenbachstrasse 1, D-85748 Garching, Germany
[167] Politecnico di Bari, Via Edoardo Orabona, 4, I-70126 Bari BA, Italy
[168] Istituto Nazionale di Fisica Nucleare, Sezione di Bari, I-70126 Bari, Italy
[169] Jacobs Technology, Inc., Huntsville, AL 35806, USA
[170] Los Alamos National Laboratory, PO Box 1663, Los Alamos, NM 87545, USA
[171] School of Physics, University College Dublin, Belfield, Stillorgan Road, Dublin 4, Ireland
[172] ISDC, Department of Astronomy, University of Geneva, Chemin d'Écogia, 16 CH-1290 Versoix, Switzerland
[173] European Space Research and Technology Centre (ESA/ESTEC), Keplerlaan 1, 2201 AZ Noordwijk, The Netherlands
[174] INAF-Institute for Space Astrophysics and Planetology, Via Fosso del Cavaliere 100, I-00133 Roma, Italy
[175] DTU Space—National Space Institute Elektrovej—Building 327 DK-2800 Kongens Lyngby Denmark
[176] Centro de Astrobiología (CAB-CSIC/INTA, ESAC Campus), Camino bajo del Castillo S/N, E-28692 Villanueva de la Cañada, Madrid, Spain
[177] Space Science Group, School of Physics, University College Dublin, Belfield, Dublin 4, Ireland
[178] IRAP, Université de Toulouse; CNRS; UPS; CNES; 9 Av. Roche, F-31028 Toulouse, France
[179] DSM/Irfu/Service d'Astrophysique, Bat. 709 Orme des Merisiers CEA Saclay, F-91191 Gif-sur-Yvette Cedex, France
[180] APC, AstroParticule et Cosmologie, Université Paris Diderot, CNRS/IN2P3, CEA/Irfu, Observatoire de Paris Sorbonne Paris Cité, France
[181] Space Research Institute of Russian Academy of Sciences, Profsoyuznaya 84/32, 117997 Moscow, Russia
[182] INAF, IASF-Milano, via E.Bassini 15, I-20133 Milano, Italy
[183] Max Planck Institute for Astrophysics, Karl-Schwarzschild-Str. 1, Garching b. Munchen D-85741, Germany